\documentclass[twocolumn, twocolappendix]{aastex631}
\usepackage{comment}
\NewPageAfterKeywords
\usepackage{graphicx}
\usepackage{url}
\usepackage{epstopdf}
\usepackage{color}
\usepackage{gensymb}
\usepackage{multirow}
\usepackage{ragged2e}
\usepackage{hyperref}
\usepackage{url}
\usepackage{amsmath} 
\usepackage{booktabs}
\usepackage{afterpage}
\usepackage{mathtools}
\usepackage{tabularx}
\usepackage{bold-extra}
\usepackage{xspace}
\usepackage{relsize}
\usepackage{braket}
\usepackage{bbm}
\usepackage{times}

\usepackage{eht}

\begin{document}

\title{Future Prospects for Constraining Black-Hole Spacetime: Horizon-scale Variability of Astrophysical Jet}

\author{Kotaro Moriyama}
\affiliation{Institut f\"ur Theoretische Physik, Goethe-Universit\"at Frankfurt, Max-von-Laue-Strasse 1, D-60438 Frankfurt am Main, Germany}
\affiliation{Mizusawa VLBI Observatory, National Astronomical Observatory of Japan, 2-12 Hoshigaoka, Mizusawa, Oshu, Iwate 023-0861, Japan}

\author{Alejandro Cruz-Osorio}
\affiliation{Instituto de Astronom\'{\i}a, Universidad Nacional Aut\'onoma de M\'exico, AP 70-264, Ciudad de M\'exico 04510, M\'exico}
\affiliation{Institut f\"ur Theoretische Physik, Goethe-Universit\"at Frankfurt, Max-von-Laue-Strasse 1, D-60438 Frankfurt am Main, Germany}

\author{Yosuke Mizuno}
\affiliation{Tsung-Dao Lee Institute, Shanghai Jiao Tong University, Shengrong Road 520, Shanghai, 201210, People's Republic of China}
\affiliation{School of Physics and Astronomy, Shanghai Jiao Tong University, 800 Dongchuan Road, Shanghai, 200240, People's Republic of China}
\affiliation{Institut f\"ur Theoretische Physik, Goethe-Universit\"at Frankfurt, Max-von-Laue-Strasse 1, D-60438 Frankfurt am Main, Germany}

\author{Christian M. Fromm}
\affiliation{Institut f\"ur Theoretische Physik und Astrophysik, Universit\"at W\"urzburg, Emil-Fischer-Str. 31, D-97074 W\"urzburg, Germany}
\affiliation{Institut f\"ur Theoretische Physik, Goethe-Universit\"at Frankfurt, Max-von-Laue-Strasse 1, D-60438 Frankfurt am Main, Germany}
\affiliation{Max-Planck-Institut f\"ur Radioastronomie, Auf dem H\"ugel 69, D-53121 Bonn, Germany}

\author{Antonios Nathanail}
\affiliation{Research Center for Astronomy, Academy of Athens, Soranou Efessiou 4, GR-115 27, Athens, Greece}
\affiliation{Institut f\"ur Theoretische Physik, Goethe-Universit\"at Frankfurt, Max-von-Laue-Strasse 1, D-60438 Frankfurt am Main, Germany}

\author{Luciano Rezzolla}
\affiliation{Institut f\"ur Theoretische Physik, Goethe-Universit\"at Frankfurt, Max-von-Laue-Strasse 1, D-60438 Frankfurt am Main, Germany}
\affiliation{Frankfurt Institute for Advanced Studies, Ruth-Moufang-Strasse 1, 60438 Frankfurt, Germany}
\affiliation{School of Mathematics, Trinity College, Dublin 2, Ireland}

\begin{abstract}
The Event Horizon Telescope (EHT) Collaboration has recently published the first horizon-scale images of the supermassive black holes {\m87} and {\sgra} and provided some first information on the physical conditions in their vicinity. 
The comparison between the observations and the three-dimensional general-relativistic magnetohydrodynamic (GRMHD) simulations has enabled the EHT to set initial constraints on the properties of these black-hole spacetimes.
However, accurately distinguishing the properties of the accretion flow from those of the spacetime, most notably, the black-hole mass and spin, remains challenging because of the degeneracies the emitted radiation suffers when varying the properties of the plasma and those of the spacetime. 
The next-generation EHT (ngEHT) observations are expected to remove some of these degeneracies by exploring the complex interplay between the disk-jet dynamics, which represents one of the most promising tools for extracting information on the black-hole spin.
By using GRMHD simulations of magnetically arrested disks (MADs) and general-relativistic radiative-transfer (GRRT) calculations of the emitted radiation, we have studied the properties of the jet and the accretion-disk dynamics on spatial scales that are comparable with the horizon. 
In this way, we are able to highlight that the radial and azimuthal dynamics of the jet are well correlated with the black-hole spin.
Based on the resolution and image reconstruction capabilities of the ngEHT observations of {\m87}, we can assess the detectability and associated uncertainty of this correlation.
Overall, our results serve to assess what are the prospects for constraining the black-hole spin with future EHT observations.
\end{abstract}



\section{Introduction}\label{sec:introduction}
The observational exploration of the physical conditions in the vicinity of a black hole is one of the most promising avenues for an accurate mapping of the spacetime and understanding of the origin of the relativistic jets.
Through groundbreaking efforts, the Event Horizon Telescope (EHT) Collaboration has recently provided images of the horizon-scale emission of the accretion flow in the vicinity of supermassive black holes in the nearby radio galaxy M87 ({\m87}) and in the Galactic center ({\sgra}).
More specifically, in 2017, the EHT array consisted of eight stations at six geographical sites that successfully observed {\m87} (\citealt{EHT_M87_PaperI, EHT_M87_PaperII,   EHT_M87_PaperIII, EHT_M87_PaperIV, EHT_M87_PaperV, EHT_M87_PaperVI}) and {\sgra} (\citealt{EHT_SgrA_PaperI, EHT_SgrA_PaperII, EHT_SgrA_PaperIII, EHT_SgrA_PaperIV, EHT_SgrA_PaperV, EHT_SgrA_PaperVI}) with the unprecedented horizon-scale resolution of $\sim 20~ \uas$. 
The imaging surveys of {\m87} and {\sgra} revealed bright emission rings that are consistent with the theoretical predictions of general relativity.

Comparing the EHT observations with the theoretical simulations enables us to constrain the black-hole spacetime within the mass and spin with general relativity as well as alternative theories.
The measured shadow diameters ($d\sim 42~\uas$ for {\m87} and $d\sim 52~\uas$ for \sgra) are consistent with the prediction of general relativity (e.g., \citealt{EHT_M87_PaperV, EHT_SgrA_PaperV, Younsi2023}) and can be also used to discriminate from general relativity's possible deviations of the black-hole spacetime (e.g., \citealt{Kocherlakota2020, EHT_M87_PaperVI, EHT_SgrA_PaperVI, Cruz2021, Vagnozzi2022}). 
The comparison of the observations with a large library of three-dimensional general-relativistic magnetohydrodynamic (GRMHD) simulations provides information on the plausible ranges of model parameters for the spacetime, the magnetic field, and the accretion flow properties (e.g., \citealt{EHT_M87_PaperV, EHT_SgrA_PaperV}).

Furthermore, the horizon-scale detection of the polarized profile around {\m87} has provided additional constraints on the magnetic fields (\citealt{EHT_M87_PaperVII, EHT_M87_PaperVIII}). 
In particular, the polarization properties of {\m87} can be explained with magnetic fields around the emission region that are predominantly poloidal, and this has allowed us to obtain the first measurements of the magnetic field strength, electron density, and temperature with horizon-scale resolution.
Moreover, by comparing the imaging results with the library of GRMHD polarized images, it was possible to confirm their consistency with magnetically arrested accretion disks (MADs).

Despite this considerable progress, with the current angular resolution of the EHT observations, it remains challenging to constrain the detailed structure of the black-hole spacetime solely from the horizon-scale images.
Despite the significant progress achieved, the current angular resolution of the EHT observations poses challenges in accurately determining the detailed structure of black-hole spacetime based solely on horizon-scale images.
For instance, according to general relativity, the diameter of the black-hole shadow is predicted to depend weakly on the observer's inclination angle and the black-hole spin, approximately $5.0\pm0.2 r_{\rm g}$, where $r_{\rm g}$ represents the gravitational radius (e.g., see \citealt{Takahashi2004, EHT_SgrA_PaperVI}).
Although these parameters are crucial in characterizing the spacetime of a black hole, achieving a more precise constraint on the black-hole parameters would require additional criteria and/or improvements in observations.
The essential role of non-thermal emission has been demonstrated with observations of {\m87} at the regime between low (43 and 86~GHz) and high (near-IR) frequencies (\citealt{Pandya2016, Davelaar2019,Cruz2022,Fromm2022}). 
Overall, GRMHD simulations have clearly demonstrated the intricate dependence of the total intensity and polarization map on the properties of both the plasma and spacetime in a complex manner.
Differentiating between them is not a straightforward endeavor and necessitates the application of innovative methodologies.

Future developments of the EHT capabilities as those explored with the next generation EHT (ngEHT) project, represent one of the most promising to address the observational and theoretical challenges discussed above.
The aim of the development is to achieve improved angular resolution across a wide range of frequencies, including 86, 230, and 345~GHz.
The possibility of improved spacetime measurements utilizing ngEHT observations is currently being intensively explored (e.g., \citealt{Chael2021, Ricarte2022}), while some research has investigated the non-thermal emission effects and constrained the observational results.
In particular, \citet{Fromm2022} have parameterized the non-thermal effect and investigated the characteristics of model images and broadband spectra spanning from the radio to the near-infrared range for  {\m87}.
The obtained best-bet models for each spin showed a strong agreement with the spectral features observed at multiple wavelengths (\citealt{Davelaar2019}).

In addition, the horizon-scale imaging of the {\m87} jet represents a promising tool for extracting information on the spacetime properties and measuring the non-thermal effects simultaneously. 
The observation of the {\m87} jet has a long history and has been conducted across a broad frequency range, from radio to $\gamma$-ray wavelengths (e.g., \citealt{Curtis1918, Reid1982, Hada2013, Kim2018a, Snios2019,  MAGIC2020}). 
The time-averaged jet morphology was further compared with 86~GHz GMVA observations of {\m87} to constrain the plausible parameter spaces of GRMHD and general-relativistic radiative transfer (GRRT) simulations (\citealt{Cruz2022,Fromm2022}).
Especially, recent 86~GHz imaging results of M87 (\citealt{Lu2023}) have revealed an edge-brightened jet that is connected to the accretion flow of the black hole. 
This observational milestone provides strong motivation for theoretical and observational studies aimed at investigating the relationship between the jet and photon ring.
The recent observational improvement and forthcoming ngEHT observations hold great potential to detect the horizon-scale structure of the jet and to capture the disk dynamics of the {\m87} through long-time monitoring (\citealt{Roelofs2023}).

In this paper, we investigate the dynamics of jets and accretion disks that can be probed through the ngEHT observations.
The previous studies in this area (e.g., \citealt{Raymond2021, Roelofs2023}) have provided evidence for the detectability of the jet base and photon ring.
We report on three-dimensional GRMHD simulations of MAD disks with the numerical code {\bhac} to obtain an accurate modeling of the accretion flow in the vicinity of the black hole (\citealt{Porth2017, Olivares2019}).
In a comparative study with other GRMHD codes utilized in the EHT Collaboration, {\bhac} demonstrated maturity, capability, and consistency in simulation results (\citealt{Porth2019}). 
Additionally, a comprehensive convergence study examining the Riemann problem, shocks, wave propagation, and accretion is available in \citet{Porth2017} and \citet{Olivares2019}.
The synthetic images of the emission are calculated with the GRRT scheme implemented in the numerical code {\bhoss}, introducing a hybrid thermal-non-thermal particle distribution function optimized by the observed continuum spectra of {\m87} (\citealt{Fromm2022, Cruz2022}).
The Radiative Transfer Scheme was rigorously tested to ensure its consistency and accuracy within the context of the general relativistic radiative transfer codes used by the EHT Collaboration (\citealt{Gold2020}).
We primarily focus on the dynamic properties of the jet and accretion disk and demonstrate that the azimuthal variation and radial jet propagation provide information on the black-hole spacetime. 
Furthermore, by comparing the simulations and reconstructed images, we measure the uncertainty and detectability of the black-hole characteristics under the planned ngEHT observations.

The plan of this paper is as follows: 
In Section~\ref{sec:methods}, we introduce the setup of the GRMHD and GRRT simulations and the strategy of the synthetic ngEHT observations. 
In Section~\ref{sec:result_fundamental}, we demonstrate the fundamental properties of the utilized GRMHD and GRRT models.
In Section~\ref{sec:imaging}, we investigate the properties of the jet-disk dynamics and demonstrate the detectability with the expected ngEHT observations.
Section~\ref{sec:velocity_origin} is instead devoted to investigating the physical origin of the detected jet dynamics discussed in Section~\ref{sec:imaging}.
Finally, in Section~\ref{sec:summary}, we present our conclusions and future prospects.


\section{Methods}
\label{sec:methods}

\subsection{General-Relativistic magnetohydrodynamic
simulations and radiative transfer}
\label{subsec:used_grmhd}

Our investigation of the dynamics of jet launching and accretion flow around a black hole has employed three-dimensional GRMHD simulations performed with the numerical code {\bhac} (\citealt{Porth2017, Olivares2019}), which solves the GRMHD equations on a Kerr background expressed in the spherical Kerr-Schild coordinates $(t, r, \theta, \phi)$.
The grid spacing is logarithmic in the radial and linear in the polar and azimuthal directions. 
Hereafter, we use geometrized units, in which the gravitational constant, $G$, and the speed of light, $c$, are set to be unity and define the gravitational radius as $r_{\rm g}:=GM/c^2$, where $M$ is the black-hole mass and the gravitational time $t_{\rm g}:=r_{\rm g}/c$.

The conditions of GRMHD simulations (the convention, resolution of the grid, and initial conditions employed in our analysis) are the same as those in previous simulations, which were used to compare with {\m87} observations (\citealt{Fromm2022, Cruz2022}). 
We take the time window with a quasi-stationary mass accretion rate between $13000\leq t/t_{\rm g}\leq 15000$, with a time resolution of $t=10~t_{\rm g} \sim 3.6~ {\rm days}$ for  {\m87}.
The scaling factors of the accretion rate (summarized in \autoref{tab:mdot}) and model parameters are selected to reproduce a consistent total flux at 230~GHz ($\sim 1~{\rm Jy}$) and a continuum spectrum density within the frequency range of $10^{10}~{\rm Hz}\leq \nu\leq 10^{16}~{\rm Hz}$.

Utilizing the GRMHD results, we calculated synthetic images using the general-relativistic ray-tracing code employed in the {\bhoss} (\citealt{Younsi2012, Younsi2020}), which solves the co-variant radiative-transfer equation. 
The calculation of the synthetic images is performed with the fast light approximation, in which the fluid is assumed not to change during the photon propagation so that an image corresponds to a single time slice.
Synchrotron radiation is estimated to be the main (only) source of emission at 230~GHz and the electron energy distribution is characterized by the so-called "kappa" model (\citealt{Xiao2006}) with a non-thermal energy contribution from the jet (\citealt{Davelaar2019}).
%
%
We focus specifically on the best-bet models of MAD with five different spins ($a_* = -0.9375, -0.5, 0.0, 0.5,$ and $0.9375$, where $a_*=J/M^2$ represents the normalized spin and $J$ denotes the angular momentum) at an inclination angle of $i=160^\circ$. 
These models exhibit a continuum spectrum consistent with the observations of {\m87} (\citealt{Cruz2022, EHT_M87_PaperV}).
The black-hole mass is set to be $6.5\times 10^9M_{\odot}$, and the distance is set to be $16.8~{\rm Mpc}$. 
Additionally, we set the azimuthal angle of the spin axis as $\phi_{\rm obs} (=\arctan(y_{\rm obs}/x_{\rm obs}))=18^\circ$ (\citealt{Fromm2022}).

\begin{table}
\caption{The mass accretion rate of the GRMHD simulations used in this study for each black-hole spin. The model names are referenced from \citet{Fromm2022}. } \centering
\begin{tabular}{lcc}
\hline
 Model   & Spin $a_*$               &Mass accretion rate $   \langle \dot{M}\rangle~{ \rm[M_{\odot}~\mathrm{yr}^{-1}]}$ \\
\hline
\hline
$\texttt{MT.M.1}$& $-0.9375$& $3.99\times 10^{-4}$ \\ 
$\texttt{MT.M.2}$& $-0.5$  & $3.42\times 10^{-4}$ \\
$\texttt{MT.M.3}$& $0$     & $2.50\times 10^{-4}$ \\
$\texttt{MT.M.4}$& $0.5$  & $2.32\times 10^{-4}$ \\
$\texttt{MT.M.5}$&$0.9375$ & $1.06\times 10^{-4}$ \\
\hline
\hline
\end{tabular}
\label{tab:mdot}
\end{table} 



\subsection{Synthetic ngEHT observations and Imaging strategy}\label{subsec:imaging}

\begin{figure}
\centering
\includegraphics[width=\linewidth]{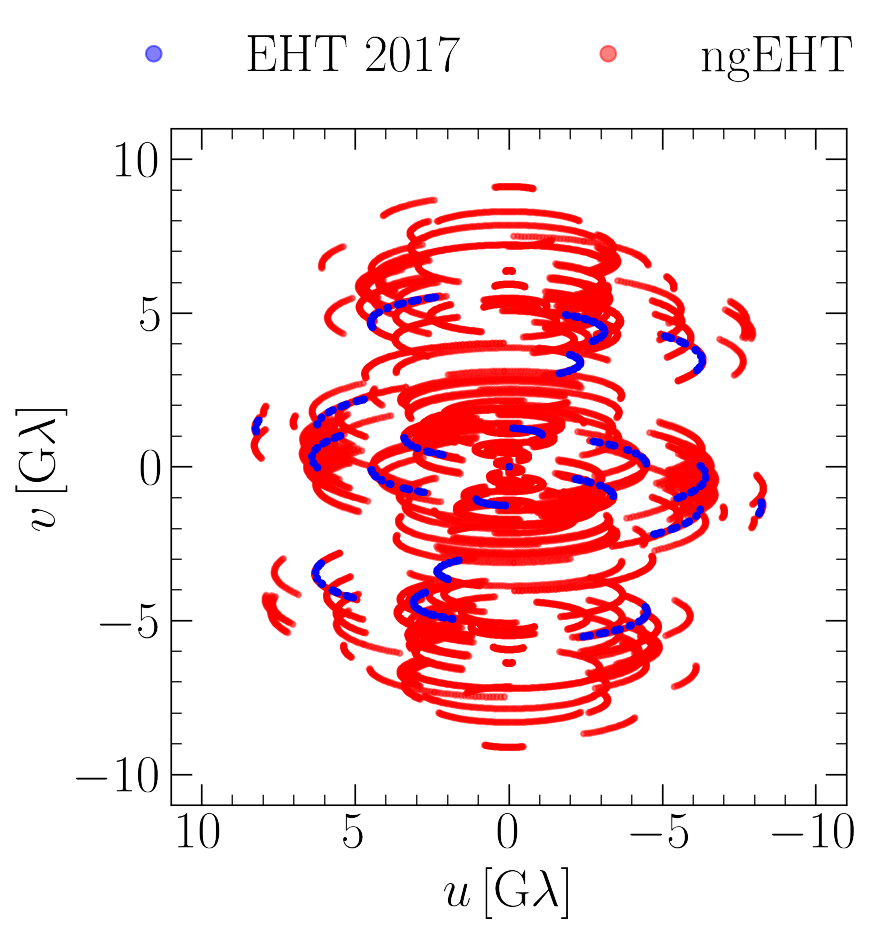}
\caption{Aggregate baseline coverage for the ngEHT observation with the expected telescopes (\citealt{Roelofs2023}). 
The blue and red data points represent the $(u,v)$ coverage obtained from the telescopes integrated in 2017 and the ngEHT, respectively, with corresponding observational time durations of 7 hours and 24 hours.
The axes of $(u, v)$ plane are in units of giga-wavelengths at the spatial frequency of 230~GHz.
}
\label{fig:visprofile}
\end{figure}

\begin{figure}
\centering
\includegraphics[width=\linewidth]{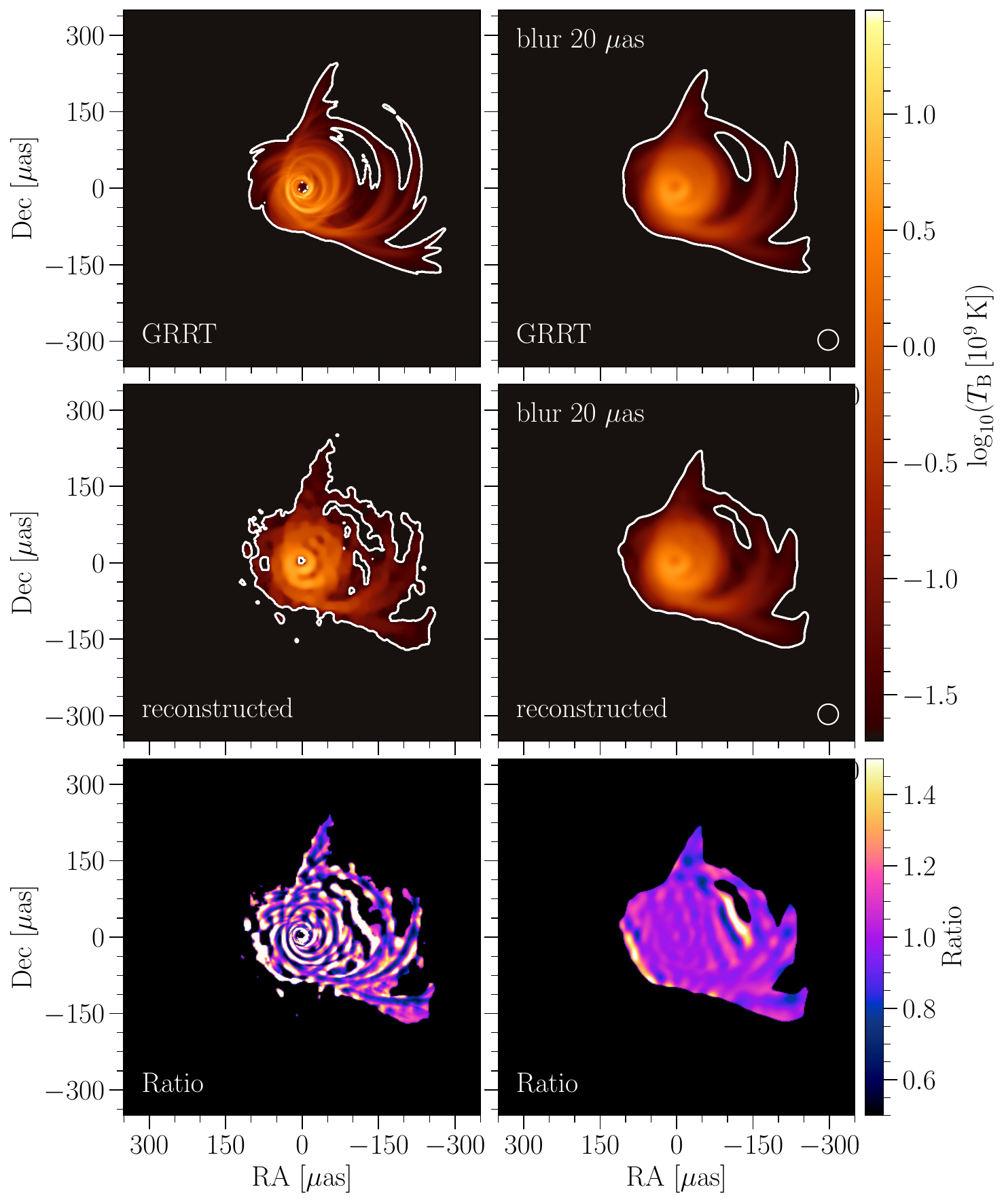}
\caption{Comparison between the top panels displaying snapshot images obtained from the GRRT simulation and the middle panels demonstrating reconstructed images using synthetic data based on the expected ngEHT observations at a frequency of $230~{\rm GHz}$ (Section~\ref{sec:imaging}), specifically focusing on the case characterized by $a_*=0.9375$.
The images in the top right and middle right panels are restored using a circular Gaussian beam with a full width at half maximum (FWHM) of $20~\uas$, as indicated in the lower right corners.
The unit represented in each image is the logarithm of the brightness temperature, denoted as $T_{\rm B}$.
The bottom panels depict the ratio between the GRRT and reconstructed images, with the bottom left panel displaying the ratio without restoration, and the bottom right panel representing the ratio after restoration with the Gaussian beam.
}
\label{fig:gt_imaging}
\end{figure}

As already anticipated, the primary aim of this paper is to explore the dynamics of the jet and of the disk, which contain information about the properties of spacetime, and to assess their detectability with the anticipated capabilities of the ngEHT.
In order to comprehensively analyze the detectable properties, it is essential to assess the impact of observational uncertainties with the expected imaging results.
The observed visibility $V(t, u, v)$ is defined as the Fourier transform of the source intensity $I(t,x_{\rm obs},y_{\rm obs})$,
\begin{equation*}
    V(t,u,v) := \int I(t,x_{\rm obs},y_{\rm obs})e^{2\pi i(ux_{\rm obs}+vy_{\rm obs})}d x_{\rm obs}d y_{\rm obs},
\end{equation*}
where $(x_{\rm obs},y_{\rm obs})$ are the Cartesian coordinates on the image plane, while $(u,v)$ are the components of the baseline vector (commonly referred to as spatial frequencies) between two antennas, which is normalized by the observing wavelength and projected on the image plane orthogonal to the source direction.
The expected telescope properties were examined in detail by \citet{Raymond2021} and \citet{Roelofs2023}, and the baseline $(u, v)$ coverage is shown in Fig.~\ref{fig:visprofile} for the observational frequency of 230~GHz. 
The measured visibility on a baseline is the summation of the true visibility and systematic noise, which is assessed with a Gaussian distribution in the complex plane with the standard deviation
\begin{equation*}
    \sigma_{i,j} = \frac{1}{\eta} \sqrt{\frac{{\rm SEFD}_i{\rm SEFD}_j}{2\Delta \nu\Delta t}}, 
\end{equation*}
where $\eta\simeq 0.88$ represents the quantization efficiency (\citealt{Thompson2017}), $\Delta \nu \simeq 8~{\rm GHz}$ refers to the bandwidth of a single spectral window for a specific band and polarization (\citealt{Roelofs2023}), and $\Delta t = 10$ s denotes the integration time (\citealt{EHT_M87_PaperIV}).
We define the system equivalent flux density (SEFD), as encompassing all thermal noises originating from the telescope receiver chains, Earth's atmosphere, and astronomical background (see \citealt{Roelofs2023} for a summary), where the subscripts $(i, j)$ refer to the telescopes involved in the respective baseline.

Assuming that the time duration of a single observation is 24 hours, during which a snapshot image does not change, we generated a data set corresponding to a time window of ($2000t_{\rm g}\sim 720~{\rm days}$).
Utilizing the open-source software \ehtim (\citealt{Chael2018}), we generate synthetic data based on the expected ngEHT $(u,v)$ coverage and source models, incorporate systematic noises into the visibilities, and perform calibration on the visibility phases.

We reconstructed the images using the open-source software {\smili}, which employs the regularized maximum likelihood (RML) method based on the synthetic data (\citealt{Akiyama2017a, Akiyama2017b}). 
For the reconstruction, we used visibilities averaged over 6 minutes, setting the imaging field of view to $1200~\uas$ and a pixel size of $4~\uas$. 
The reconstruction process was initiated using an image characterized by a circular Gaussian profile with a FWHM of $50~\uas$.
Furthermore, we utilized imaging regularizations of Total Variation (TV), with a parameter value of $10^6$, and relative entropy, with a value of $1$ (\citealt{EHT_M87_PaperIV}).

We present an example of GRRT (top panels) and reconstructed images (middle) in Fig.~\ref{fig:gt_imaging}.
Each image in the top-right and middle-right panels has been convolved with a circular Gaussian function corresponding to the typical resolution of the EHT ($20~\uas$).
Note that the reconstructed image accurately captures the extended-jet features, while also exhibiting a certain level of noise structure, which can be attributed to systematic errors and the constrained $(u,v)$ coverage.
The intensity ratio between the GRRT and reconstructed images is shown in the bottom panels.
The standard deviation of the intensity ratio for the unconvolved images ($\sim 2.8$) is mitigated through convolution at the nominal resolution ($\sim 0.1$).
In Section~\ref{sec:imaging}, we use the convolved images to investigate the common dynamic properties consistent between GRRT and reconstructed images.


\section{Results: fundamental properties of GRRT images}\label{sec:result_fundamental}

\subsection{Time-averaged image properties}\label{subsec:avgimage}
\begin{figure*}[t!]
\centering
\includegraphics[width=\linewidth]{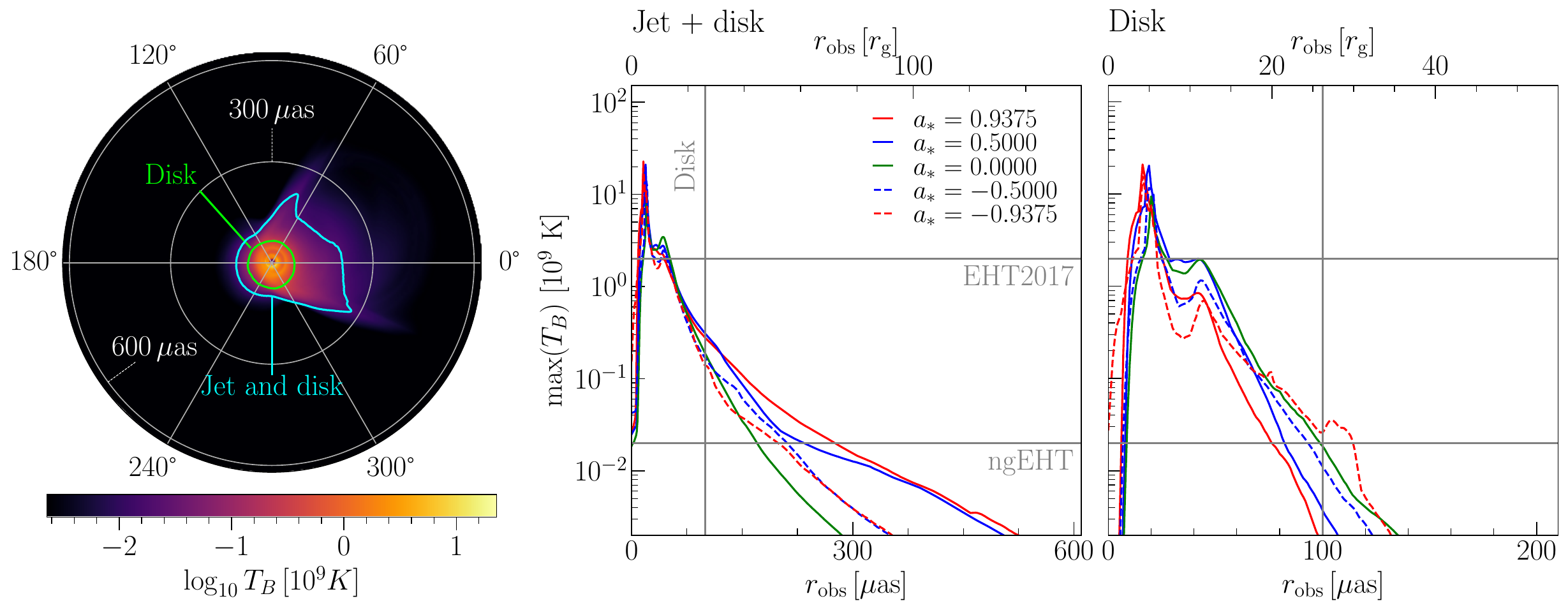}
\caption{
The time-averaged properties of the accretion disk and the jet at $230~{\rm GHz}$.
The left panel displays the time-averaged GRRT image in spherical polar coordinates for the case of a spin and inclination angle of $(a_*, i)=(0.9375, 160^\circ)$.
The green and cyan curves represent the disk and jet$+$disk regions, respectively, which can be detected with the typical ngEHT sensitivity ($=2\times 10^7~{\rm K}$).
In the middle panel, the radial profile of the azimuthal maximum brightness temperature of the jet$+$disk is shown.
The expected sensitivity of the 2017 EHT and ngEHT observations is presented as horizontal gray lines.
In the right panel, the same format is followed as in the middle panel, but for the disk emission defined by the Bernoulli parameter (${\rm Be}<1.02$).
The vertical gray lines in the middle and right panels indicate a radius of $100~{\rm \uas}$, which is the typical radius of the disk region for each spin case.
}
\label{fig:avg_jet}
\end{figure*}

\begin{table}
\caption{
Time average and standard deviation of jet$+$disk and disk sizes using the ngEHT sensitivity as defined in Fig.~\ref{fig:avg_jet}, where $a_*$ represents the dimensionless black-hole spin, and $r_{\rm obs,, jet+disk}$ and $r_{\rm obs,,disk}$ denote the radial sizes of the jet$+$disk and disk regions, respectively. 
} \centering
\begin{tabular}{ccc}
\hline
$a_*$ & $r_{\rm obs,~ jet+disk}~ [{\rm \uas}]$ & $r_{\rm  obs,~disk}~ [{\rm \uas}]$\\
\hline
~0.9375  & $287\pm 33$  & $81\pm 12$\\
~0.50  & $251\pm 20$  & $96\pm 7$\\
~0.00  & $181\pm 15$  & $110\pm 10$\\
-0.50  & $227\pm 25$  & $101\pm 19$\\
-0.9375  & $223\pm 41$  & $115\pm 18$\\
\hline
\end{tabular}
\label{tab:jet_disk_region}
\end{table} 

Identifying the spatial regions on the image plane corresponding to the jet and the disk is essential for extracting physical information from both the GRRT and reconstructed images. 
We calculated the radiation emitted from each region, the boundary of which is defined by the Bernoulli parameter ${\rm Be}:=-hu_t=1.02$, where $h$ is the specific enthalpy and $u_t$ is the time component of the covariant four-velocity (\citealt{Fromm2022}). 
Along each ray trajectory, we set the emissivity to zero outside the corresponding region while maintaining the absorptivity to the relevant value.

To demonstrate the essence of our approach, we show in Fig.~\ref{fig:avg_jet} the time-averaged properties of the jet and the disk structures as averaged over the entire simulation (i.e., from $t/t_{\rm g}=0$ to $2000$).
The left panel shows the average of the synthetic images for a Kerr black hole with $a_*=0.9375$ with the green and cyan curves identifying the detectable regions of the disk and jet+disk using the typical ngEHT sensitivity ($=2\times 10^7$ K, \citealt{Doeleman2019, Raymond2021, Chael2021}). 
The disk radiation concentrates within the radial region of the image plane at approximately $r_{\rm obs} =\sqrt{x_{\rm obs}^2+y_{\rm obs}^2}\sim 100~\uas \simeq 26 r_{\rm g}$, while the jet radiation extends to the radial region $100~\uas\leq r_{\rm obs}<300~\uas$.

Each curve in the middle panel of Fig.~\ref{fig:avg_jet} represents the radial profile of the azimuthally maximum brightness temperature, where the black, blue, and red curves correspond to $a_*= 0$, $|a_*|=0.5$, and $|a_*|=0.9375$, respectively. 
For $a_*> 0$ (solid curve), the radial extent of the jet increases with $|a_*|$: the jet extends towards $r_{\rm obs}\sim 500~\uas$ ($300~\uas$) with a dynamic range of $10^{-4}$ in the spinning (non-spinning) black hole cases (solid curves).
In the case of $a_*< 0$ (dashed curves), the radial extent of the jet is similar with a dynamic range of $10^{-4}$. 
The existence of a powerful jet in the case of a prograde black hole was also reported by other studies (e.g., \citealt{Blandford1977, Tchekhovskoy2012, Fromm2022, Narayan2022}).

The right panel of Fig.~\ref{fig:avg_jet} shows the radial profile of the time-averaged disk emission. 
The profile highlights the photon-ring region, with fainter components at the radial center and peak components at $20~\uas =5.2~r_{\rm g}$. 
These features indicate the black-hole shadow and the photon ring, respectively.
The disk extends radially from $20~\uas$ to $100~\uas$, and the brightness temperature within the radial region of $r_{\rm obs}\sim 100 ~ r_{\rm g}$ decreases.
The disk extends radially from $20~\uas$ to $100~\uas$, and the brightness temperature within the radial region of $r_{\rm obs}\sim 100 ~ r_{\rm g}$ decreases outwards.

The detectable regions of the disk and jet components can be assessed using the sensitivities of the 2017 EHT and the expected ngEHT observations, as represented by the gray horizontal lines in the middle and right panels. 
The emission from the disk and the jet regions can be detected within the ngEHT sensitivity at radii of $r_{\rm obs}<100~\uas$ and $100~\uas \lesssim r_{\rm obs} \lesssim 300~ \uas$ respectively. 
This would mark a significant improvement compared to the capabilities of the 2017 EHT observations. 
The uncertainties related to time variation are summarized in \autoref{tab:jet_disk_region}. 
In light of Fig.~\ref{fig:avg_jet} and \autoref{tab:jet_disk_region}, we define the radii of the disk and the jet regions on the image plane as $r_{\rm obs}<100~\uas$ and $100~\uas \leq r_{\rm obs}$, respectively.



\subsection{Light-curve properties}\label{subsec:lightcurve}

\begin{figure*}
\centering
\includegraphics[width=\textwidth]{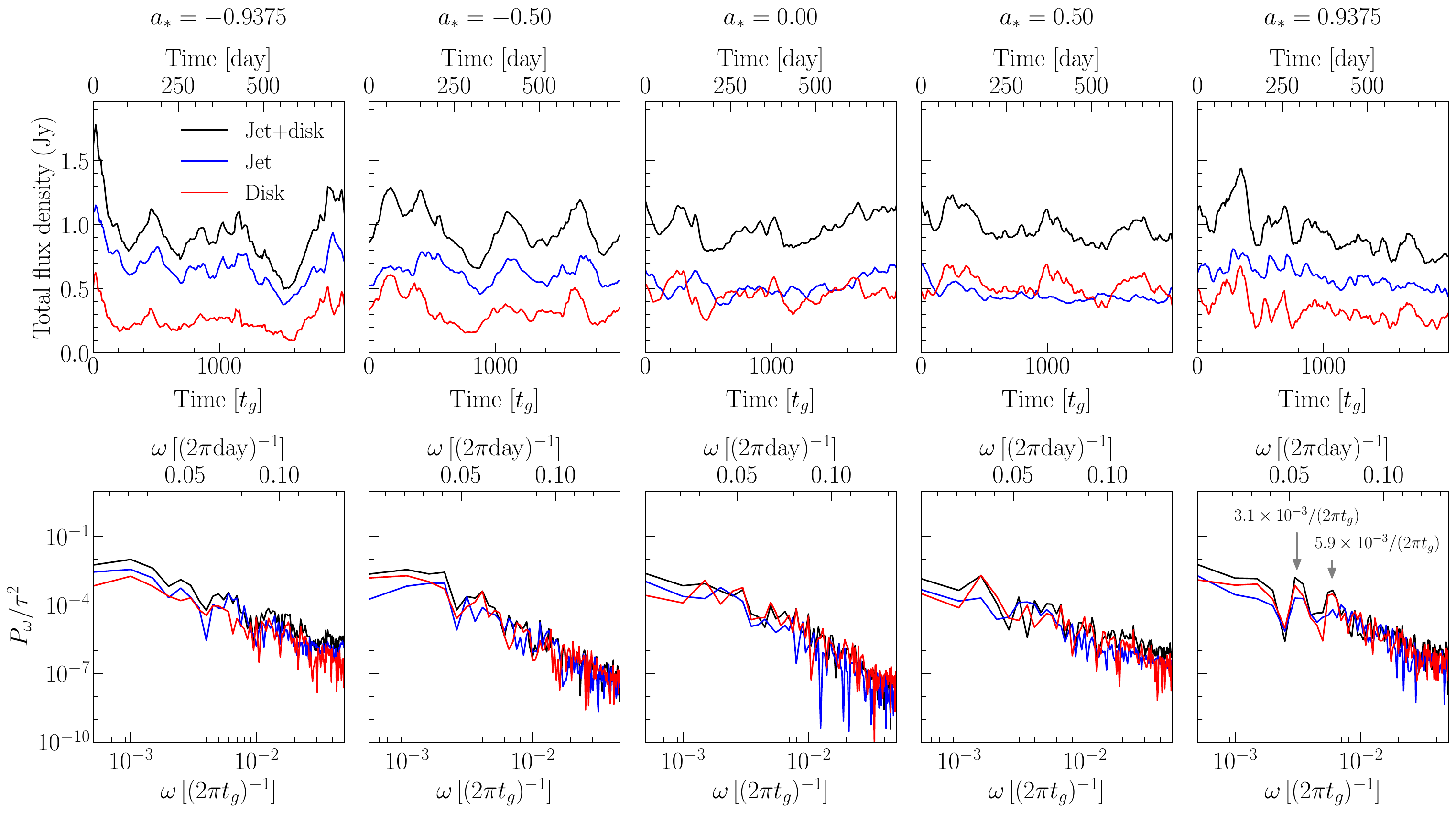}
\caption{
The spin dependency of the $230~{ \rm GHz}$ light-curves and power spectral densities (PSDs).
From left to right, the different panels refer to black holes with $a_*=-0.9375$ to $0.9375$.
Top panels: the light curves emitted from both jet and disk (black), jet-only (blue), and disk-only (red) regions. 
Bottom panels: the power spectral densities of the light curves from the upper panels, emitted from each region, where $\tau=2000t_{\rm g}$ is the entire simulation time window.
}
\label{fig:lc_psd}
\end{figure*}

\begin{table}
\caption{
Summary of the quasi-periodic oscillation (QPO). 
Here, $\omega$ represents the peak frequency detected in the power spectral densities (as seen in Fig.~\ref{fig:lc_psd}), $Q$ denotes the quality factor of the QPO, and $\Delta \omega$ corresponds to the full width at half-maximum (FWHM) of the peak.
}
\centering
\begin{tabular}{cccc}
\hline
$a_*$ & QPO frequency $\omega$ [$2\pi t_{\rm g}^{-1}$] & $1/\omega$ [$t_{\rm g}$]& $Q=\omega/\Delta \omega$ \\
\hline
~0.9375  & $3.1\times 10^{-3},$ $5.9\times 10^{-3}$ & $320,$ $170$  & 8, 2\\
~0.50    &  - & -  & -\\
~0.00    &  - & -  & -\\
-0.50    & $1.9\times 10^{-3}$ & $516$  & 6\\
-0.9375  & - & - & - \\
\hline
\end{tabular}
\label{tab:qpo}
\end{table} 

We report in Fig.~\ref{fig:lc_psd} the temporal variability of the emission from all regions, including the jet and the disk.
The top panels of Fig.~\ref{fig:lc_psd} show the light curves for different black-hole spins ($a_* = -0.9375, -0.5, 0.0, 0.5,$ and $0.9375$), as emitted from all regions (black curve), from the jet (blue) and from the disk (red). 
The light curves exhibit variability within the entire simulation time window, defined as $\tau = 2000t_{\rm g}$. 
The mean and standard deviation of the total jet$+$disk light curve are comparable across each spin case ($\sim 0.96 \pm 0.16~{\rm Jy}$) and are consistent with the core flux of {\m87} observed in previous EHT observations ($\sim 1.1-1.2~{\rm Jy}$, \citealt{Doeleman2012, EHT_M87_PaperIV}).

Most models demonstrate a dominance of jet radiation over that from the disk. 
With a ratio of the jet flux to the disk flux that is $\langle F_{\rm jet}\rangle_t/\langle F_{\rm disk}\rangle_t=2.6, 1.9, 1.1, 0.9,$ and $1.8$ for $a_*=-0.9375, -0.5, 0.0, 0.5,$ and $0.9375$, respectively, where the brackets $\langle~\rangle_t$ represent the time average. 
For both $a_*\geq 0.5$ and $a_*<0$ cases, the ratio monotonically increases with the absolute value of the spin, indicating the generation of a powerful jet for rapidly rotating black holes.

Our best-bet GRMHD models have smaller fluxes in the mid-plane region around the black hole (with a field of view of approximately $100~\uas$ and a polar angle range of $57.3^\circ\leq \theta\leq 122.7^\circ$) compared to those of the thermal models based on the EHT GRMHD library (\citealt{EHT_M87_PaperV}). 
To investigate this difference, we calculated the ratio of the total flux in the mid-plane to that in all regions, utilizing an $80~\uas$ field of view, in line with the analysis in Appendix B of \citet{EHT_M87_PaperV}. 
The ratios of the total flux density are $66~\%,~ 85~\%,~ 79~\%,~ 87~\%,$ and $81~\%$ for $a_*=-0.9375, -0.5, 0.0, 0.5,$ and $0.9375$, respectively, whereas the MAD thermal models yield a value of about $90~\%$ (see Appendix B in \citealt{EHT_M87_PaperV}). 
This discrepancy can be attributed to the impact of non-thermal particles, which result in a comparable continuum spectrum spanning from the radio to the near-infrared range and influence the jet morphology at a frequency of $86~{\rm GHz}$ in the case of {\m87} (see \citealt{Fromm2022, Cruz2022}).

The variability characteristics of each light curve are also apparent in the power spectral densities (PSDs) $P_{\omega}$, where $\omega$ denotes the angular frequency (bottom panels of Fig.~\ref{fig:lc_psd}).
The frequency profiles of the PSDs, that are characterized by $P_{\omega}\propto \omega^{-2.4\pm 0.5}$, are broadly consistent with those produced by red noise ($P_{\omega}\propto \omega^{-2}$). 
This property was already reported in previous research that utilized a large library of GRMHD simulation models without non-thermal particles (\citealt{Georgiev2022}).

Interestingly, some PSDs display a feature that could be associated with a quasi-periodic oscillation (QPO). 
This trend is especially pronounced for $a_*=0.9375$, where the PSD exhibits peak components at time scales of $1/\omega \sim 320t_{\rm g}$ and $170t_{\rm g}$ with corresponding quality factors of $\omega/\Delta \omega\sim 8$ and $2$, respectively. 
Here, $\Delta \omega$ represents the full width at half maximum (FWHM), determined by fitting the peaks with the Lorentzian function. 
This pattern is reminiscent of QPO phenomena observed in some stellar-mass and supermassive black holes and could be employed to deduce the properties of the background spacetime (e.g., \citealt{Abramowicz2003, Rezzolla2003, Reynolds2009, Kato2008, McKinney2012, Smith2021}). 
We intend to undertake a detailed investigation of the QPOs in future works.



\section{Results: movie morphologies}\label{sec:imaging}
In what follows we discuss the analysis of the dynamics of the disk and of the jet by employing synthetic images generated from GRRT simulations.
Our assessment of the observational detectability and uncertainty has been achieved through the comprehensive imaging analysis of the expected ngEHT observations, which we summarized in Section~\ref{subsec:imaging}.
Our goal is to demonstrate that the structures of the jet and of the disk exhibit azimuthal and radial fluctuations that depend on the black-hole spin. 


\subsection{Variability of the Azimuthal Distribution}\label{subsec:structure_variation}

\begin{figure*}
\centering
\includegraphics[width=\linewidth]{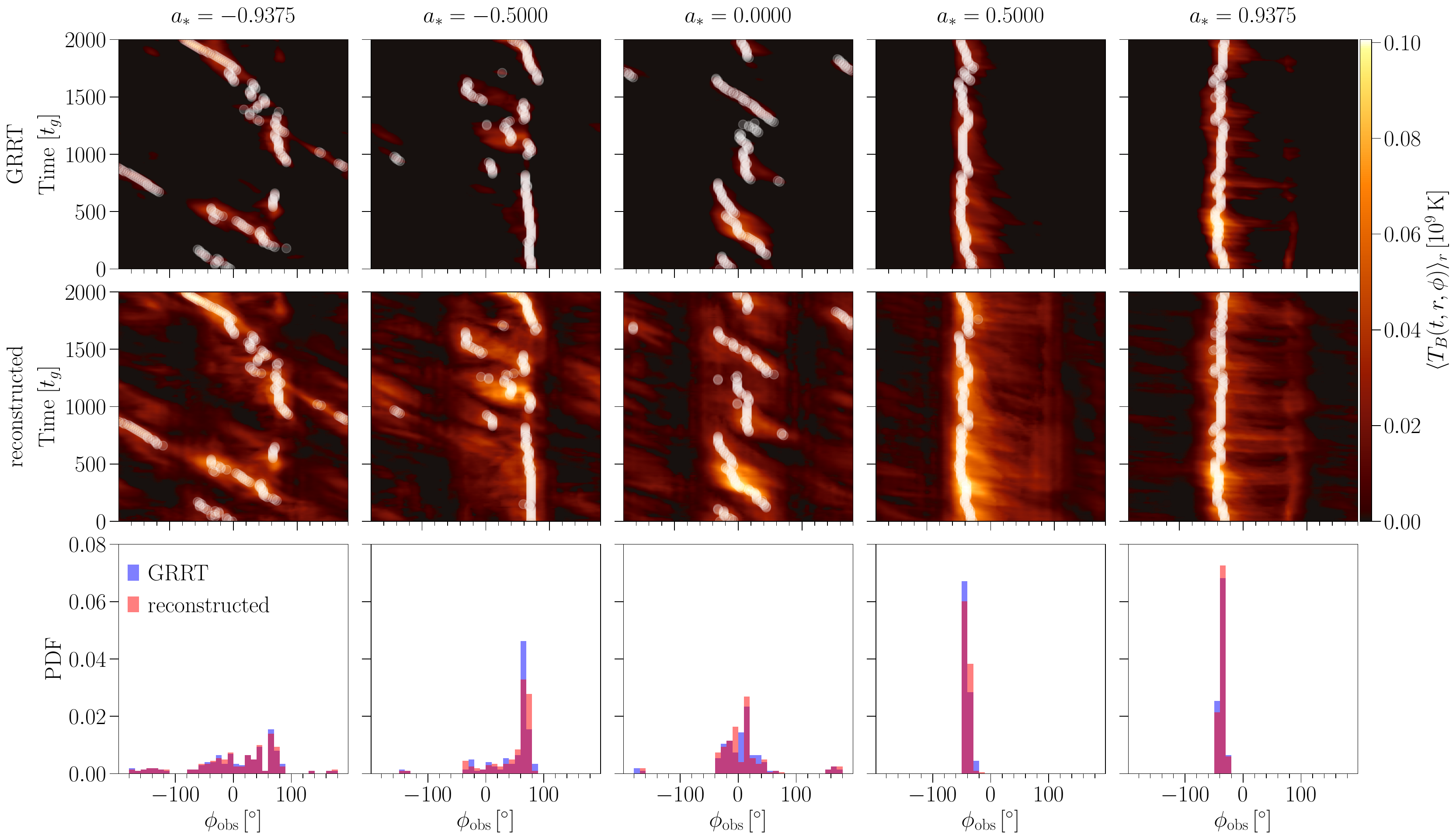}
\caption{
Azimuthal profiles of the image brightness radially averaged within the jet region ($100~\uas\leq r_{\rm obs}<300~\uas$).
From left to right, the different panels refer to black holes with $a_*=-0.9375$ to $0.9375$.
The top and middle panels depict the brightness temperature on the plane of the azimuthal angle and time $(\phi, t)$, utilizing the GRRT and reconstructed images, respectively. 
The white points denote the maximum azimuthal brightness at each time. 
The bottom panels exhibit the distributions of the peak azimuthal angles obtained from the upper panels. 
The blue and red histograms represent the estimates derived from the GRRT and reconstructed images, respectively.
}
\label{fig:phi_t_Tb_jet}
\end{figure*}

\begin{figure*}
\centering
\includegraphics[width=\linewidth]{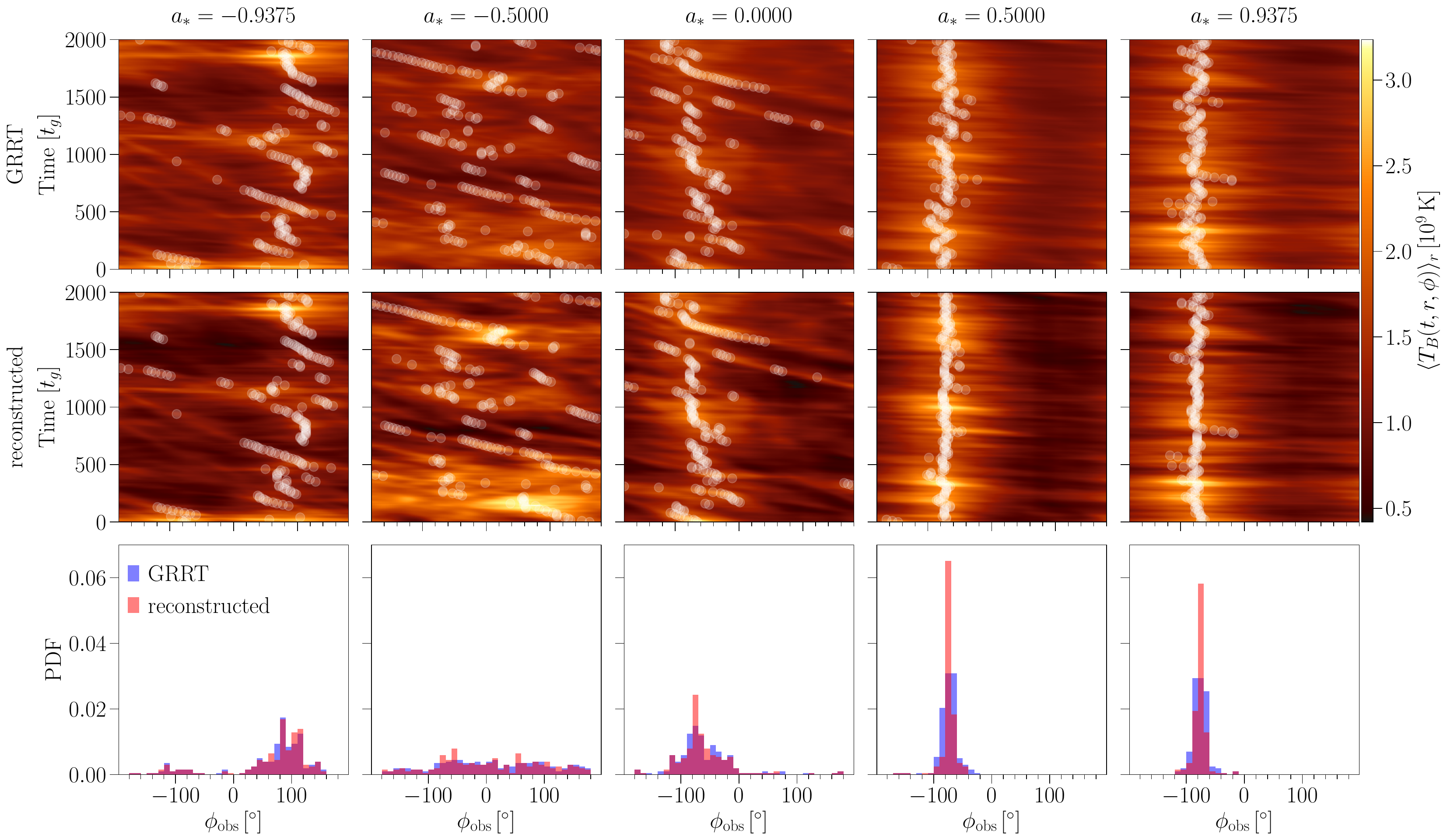}
\caption{
The same as in Fig.~\ref{fig:phi_t_Tb_jet}, but each profile was radially averaged only in the disk region $r_{\rm obs}<100~\uas$. 
}
\label{fig:phi_t_Tb_disk}
\end{figure*}

\begin{figure}
\centering
\includegraphics[width=\linewidth]{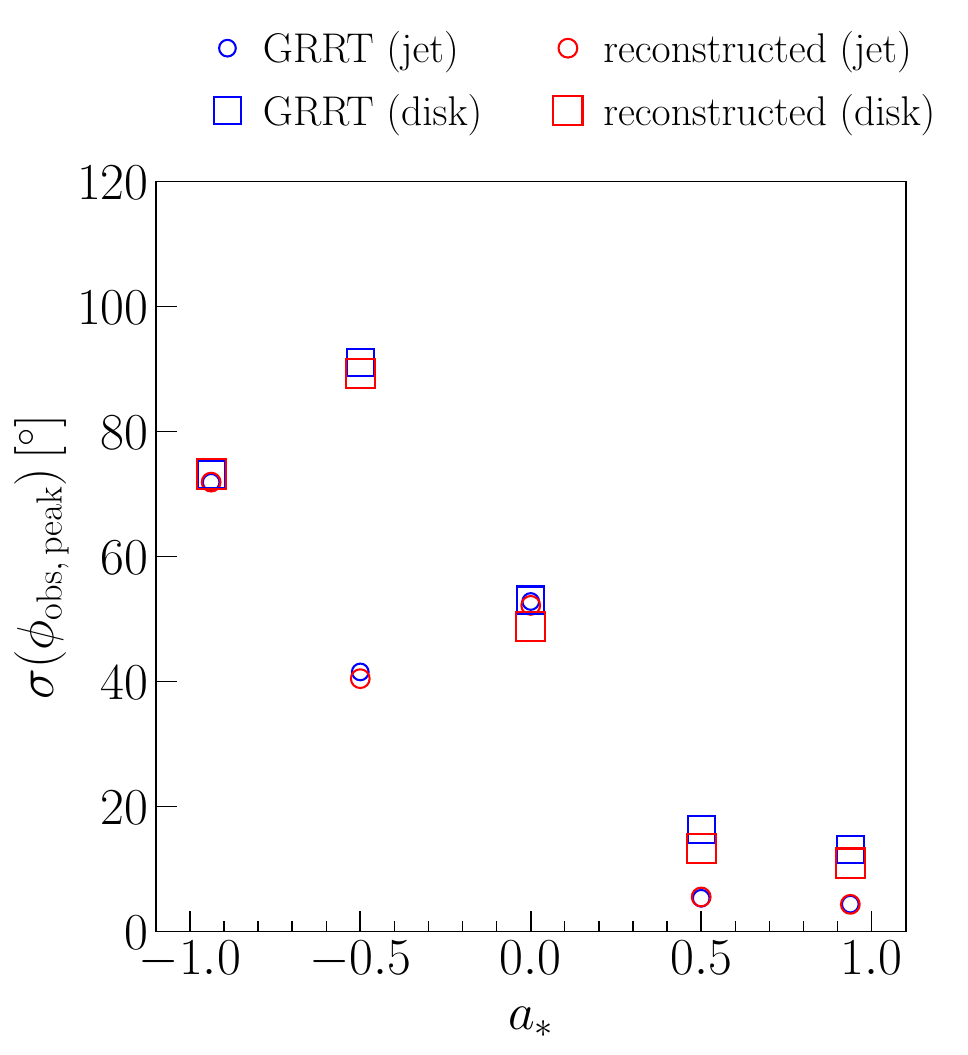}
\caption{
The spin dependency of the standard deviation of the peak azimuthal angles in the jet (represented by circles) and the disk (squares) regions. 
The blue and red markers represent the estimated values obtained from the GRRT and reconstructed images, respectively.
}
\label{fig:abh_phi_peak}
\end{figure}

In Fig.~\ref{fig:phi_t_Tb_jet}, we show the azimuthal profiles of the jet, where each image is radially averaged within the jet region; i.e., $100~\uas\leq r_{\rm obs}<300~\uas$ (corresponding to $26\leq r_{\rm obs}/r_{\rm g}< 79$). 
The azimuthal profile corresponding to a negative spin exhibits substantial variability along the temporal direction, while profiles with positive spins display relatively stable behavior over time. 
The white points $\phi_{\rm obs,~ peak}(t)$ illustrate the temporal evolution of the azimuthal angle at maximum brightness.
In the instance of $a_* =-0.9375$, intermittent azimuthal rotations of $\phi_{\rm obs,~ peak}(t)$ occur in the same direction as the accretion flow (clockwise) with a yearly timescale ($\sim 1000t_{\rm g}\simeq1.0~\rm{year}$ for $M=6.5\times 10^9M_{\odot}$). 
For example, the white points within the range of $-180^\circ\leq \phi_{\rm obs} \leq 60^\circ$ and $170 \leq t/t_{\rm g}\leq 880$ exhibit an angular velocity of $d\phi_{\rm obs}/dt\sim -0.4^\circ/t_{\rm g}$.
The substantial variation in the azimuthal angle on a yearly scale, $\delta \phi_{\rm obs}\sim 240^\circ$, is a distinctive characteristic of the case with $a_*=-0.9375$, separating it from the jets produced by black holes with $-0.5\leq a_*\leq 0$ and much more evidently from black hole with $a_*=0.9375$.

The azimuthal profiles derived from the reconstructed images shown in the middle panels of Fig.~\ref{fig:phi_t_Tb_jet} clearly highlight that the azimuthal variability is observationally detectable.
The comparison between the top and middle panels reveals consistent time development of $\phi_{\rm obs,~peak}(t)$ in both the GRRT and reconstructed images. 
The time-aggregated distributions of $\phi_{\rm obs~peak}(t)$ are reported in the bottom panels using GRRT (blue histogram) and reconstructed images (red histogram).
Note how the negative spin cases exhibit a wide range of peak azimuthal angles, whereas the positive spin cases display sharply peaked histograms.

Figure~\ref{fig:phi_t_Tb_disk} is similar to Fig.~\ref{fig:phi_t_Tb_jet}, but in it we show the azimuthal variability in the disk region only, $r_{\rm obs}<100~\uas$  (corresponding to $r_{\rm obs}/r_{\rm g}< 26$). 
The temporal evolutions of $\phi_{\rm obs,~ peak}(t)$ (white points) show more substantial variability than those in the jet region (Fig.~\ref{fig:phi_t_Tb_jet}) and a similar trend is evident in the jet region, where the variations of the white points are larger for the negative spin cases compared to the positive spin cases.

In Fig.~\ref{fig:abh_phi_peak}, we summarize the standard deviation of the azimuthal angle at peak brightness $\phi_{\rm obs,~ peak}(t)$ in the jet (circle points) and the disk regions (square points), utilizing both GRRT (blue) and reconstructed images (red).
The peak azimuthal angle of the disk region is determined by the gravitational lensing and beaming effects, as well as the brightness distribution of the accretion flow.
The transition value between $a_*\leq 0$ and $a_*>0$ is approximately $\sigma(\phi_{\rm obs,~ peak})\sim 30^\circ$ in both regions, which allows us to constrain the direction of the black-hole rotation.
The standard deviation of the peak azimuthal angles exhibits similar values across nearly all spin cases in both the jet and the disk images.

Moreover, we can see that the disk component $\sigma (\phi_{\rm obs, peak})\sim 90^\circ$ has an azimuthal variation that is larger than that of the jet component ($\sim 40^\circ$) in the case of $a_*=-0.5$, which can be explained with the different rotational direction between the black hole and accretion flow.
For $a_* \geq 0$, the gravitational lensing and beaming effects cause the mean peak position angle to be distributed around $\sim -60^\circ$ (see Fig.~\ref{fig:phi_t_Tb_disk}).
On the other hand, at $a_* = -0.9375$, the mean peak azimuthal angle is distributed around $\sim 63^\circ$ due to the strong frame dragging effect of the counter-clockwise rotation of the black hole, which disrupts the rotation direction of the accretion flow in the opposite direction.
In the case of $a_* = -0.5$, the effect of the black hole's frame dragging is not as dominant over the opposing flow of the accretion disk as it is when $a_* = -0.9375$, and thus, large fluctuations occur due to the competition between these two effects.
This unique property of azimuthal variation seen in the case of $a_*=-0.5$ has potential utility in distinguishing between observational results for $a_*=-0.9375$, $-0.5$, and $0.0$.



\subsection{Radial wave propagation of jet inhomogeneities}\label{subsec:radial_wave}

\begin{figure*}
\centering
\includegraphics[width=\linewidth]{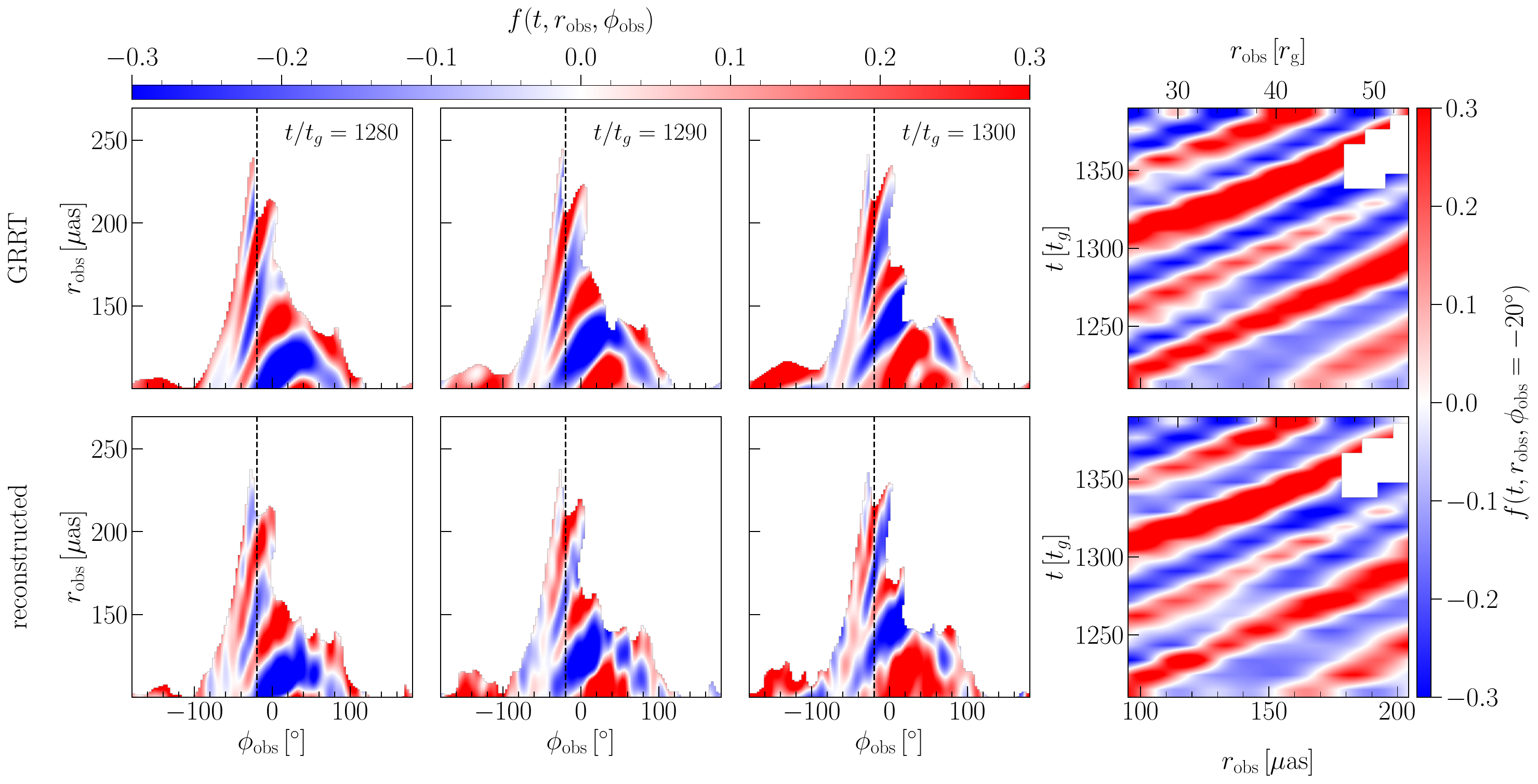}
\caption{
The radial propagation of wave inhomogeneity within the jet region in the case of $a_*=0.9375$.
The top and bottom panels represent the distributions using the GRRT and reconstructed images, respectively.
The three panels on the left-hand side show the polar distribution of temperature inhomogeneity at corresponding timestamps ($t/t_{\rm g} = 1280,~1290,$ and $1300$).
The panels on the right-hand side display spacetime diagrams illustrating the inhomogeneity distributions at an azimuthal angle of $\phi_{\rm obs} = -20^\circ$.
}
\label{fig:demo_spacetime_diagram}
\end{figure*}

In this section, we shift our focus to the radial variability in the jet region and propose observational quantities to estimate the magnitude of the black-hole spin.
To this scope, we investigate the time development of radial jet propagation. We introduced a distribution of the relative brightness temperature
\begin{eqnarray*}
f(t,r_{\rm obs},\phi_{\rm obs}) &:=&{ T_{B}(t+\Delta t, r_{\rm obs},\phi_{\rm obs})\over T_{B}(t, r_{\rm obs},\phi_{\rm obs})}-1,
\end{eqnarray*}
where $T_B (t, r_{\rm obs}, \phi_{\rm obs})$ represents the brightness temperature in polar coordinates at the observational time $t$, and $\Delta t=10t_{\rm g}$ denotes the time interval of the GRRT simulations.
We calculated the relative brightness in regions where $T_B (t, r_{\rm obs}, \phi_{\rm obs})$ and $T_B (t+\Delta t, r_{\rm obs}, \phi_{\rm obs})$ exceed the sensitivity threshold of the ngEHT, $T_{B,{\rm thres}} =2\times 10^7~{\rm K}$.

The GRRT images clearly show the propagation of inhomogeneous wave-like features in the radial direction.
In Fig.~\ref{fig:demo_spacetime_diagram}, we report the inhomogeneous radial wave observed in the case of $a_* =0.9375$.
The relative distributions between the neighboring time stamps (top and left three panels) demonstrate the propagation of inhomogeneity in the radial direction.
Comparing the relative brightness distributions between the GRRT (top left panels) and the reconstructed images (bottom left panels) provides evidence that similar wave-like features propagating outwards in the radial direction can be detected through expected ngEHT observations.

To better highlight the temporal evolution of the radial wave and measure its speed, we show in the right panels of Fig.~\ref{fig:demo_spacetime_diagram} the time development of the radial profile at the azimuthal angle of $\phi_{\rm obs}=-20^\circ$.
Intermittent wave propagation is clearly observed, as evidenced by the presence of five red curves in the spacetime diagram within the time range of $1200\leq t/t_{\rm g} < 1400$ based on both the GRRT (top panel) and reconstructed images (bottom panel).
Each red/blue stripe exhibits a linear behavior, indicating the intermittent radial propagation of wave inhomogeneity with an almost continuous velocity.
This feature sets the stage for our forthcoming examination of the spin dependence on the radial wave velocity in the subsequent section.



\subsection{Radial velocity of jet inhomogeneity}\label{subsec:spin_vr}

\begin{figure*}
\centering
\includegraphics[width=\linewidth]{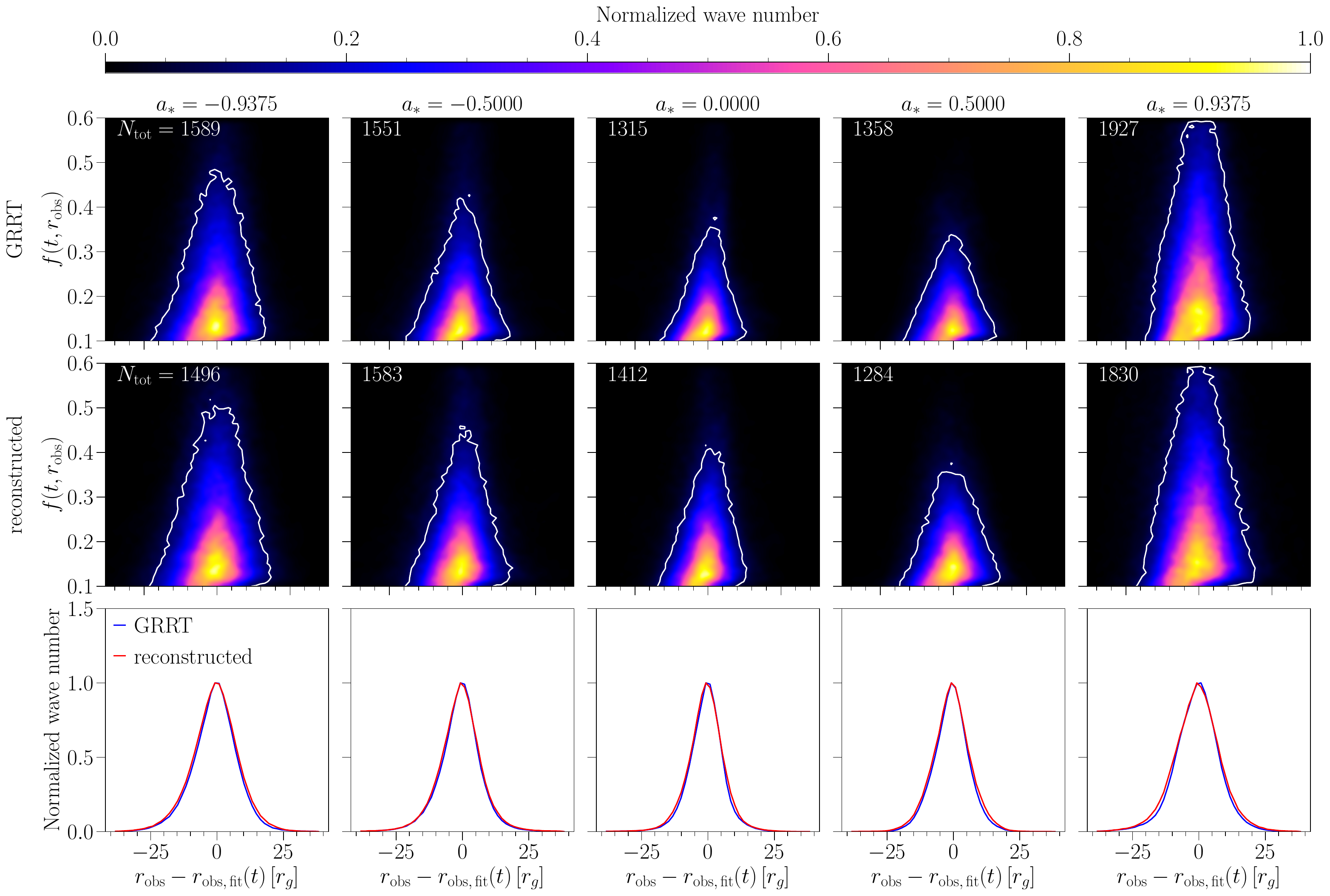}
\caption{
The number of detected waves with respect to the relative brightness $f$ across the entire time and azimuthal domains (from left to right: $a_*=-0.9375, -0.5, 0.0, 0.5,$ and $0.9375$).
The top and middle panels show the distributions of $N/{\rm max}(N)$ at the residual radius $r_{\rm obs}-r_{\rm obs,~fit}$ and the relative brightness of each wave, utilizing the GRRT and reconstructed images, respectively.
Here, $r_{\rm obs,~fit}$ represents the fitting radius of each wave in the spacetime diagram, the white curve represents the boundary where $N/{\rm max}(N)=0.1$, and the leftmost number $N_{\rm tot}$ denotes the total number of detected waves.
The bottom panels depict the vertical integration of $N$ from the upper panels, with normalization to its maximum value.
}
\label{fig:res_fit}
\end{figure*}

\begin{figure*}
\centering
\includegraphics[width=\linewidth]{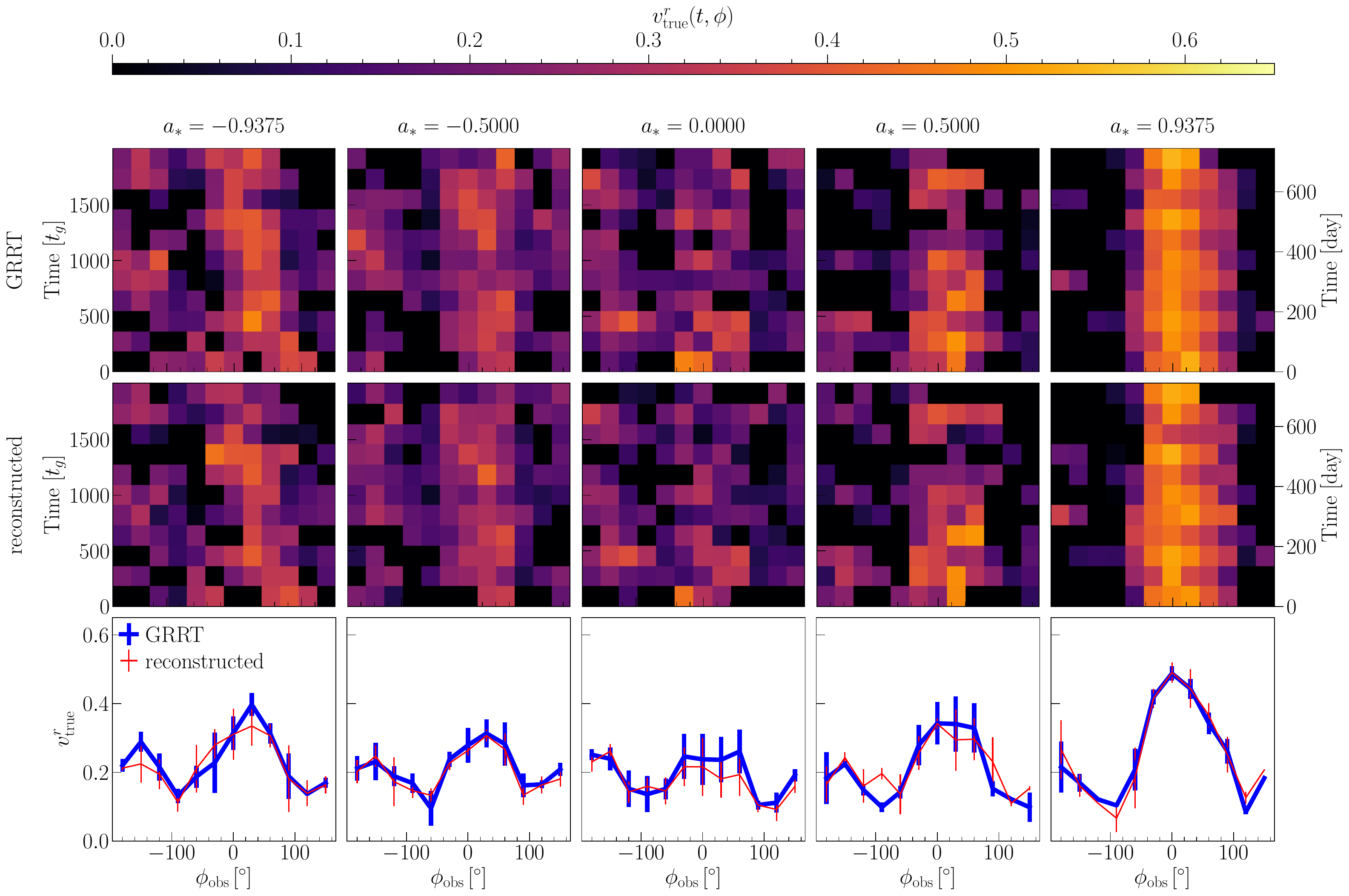}
\caption{
Azimuthal and temporal distributions of the radial wave velocities $v^r_{\rm true}$ for each spin.
The top and middle panels show the distribution of radial velocity using the GRRT and reconstructed images, respectively.
The lower panels present the median values of the radial velocity in the temporal direction, derived from both the GRRT (blue) and reconstructed (red) images, with the error bars indicating the corresponding median absolute deviations.
}
\label{fig:phi_t_a_vr}
\end{figure*}

\begin{figure}
\centering
\includegraphics[width=\linewidth]{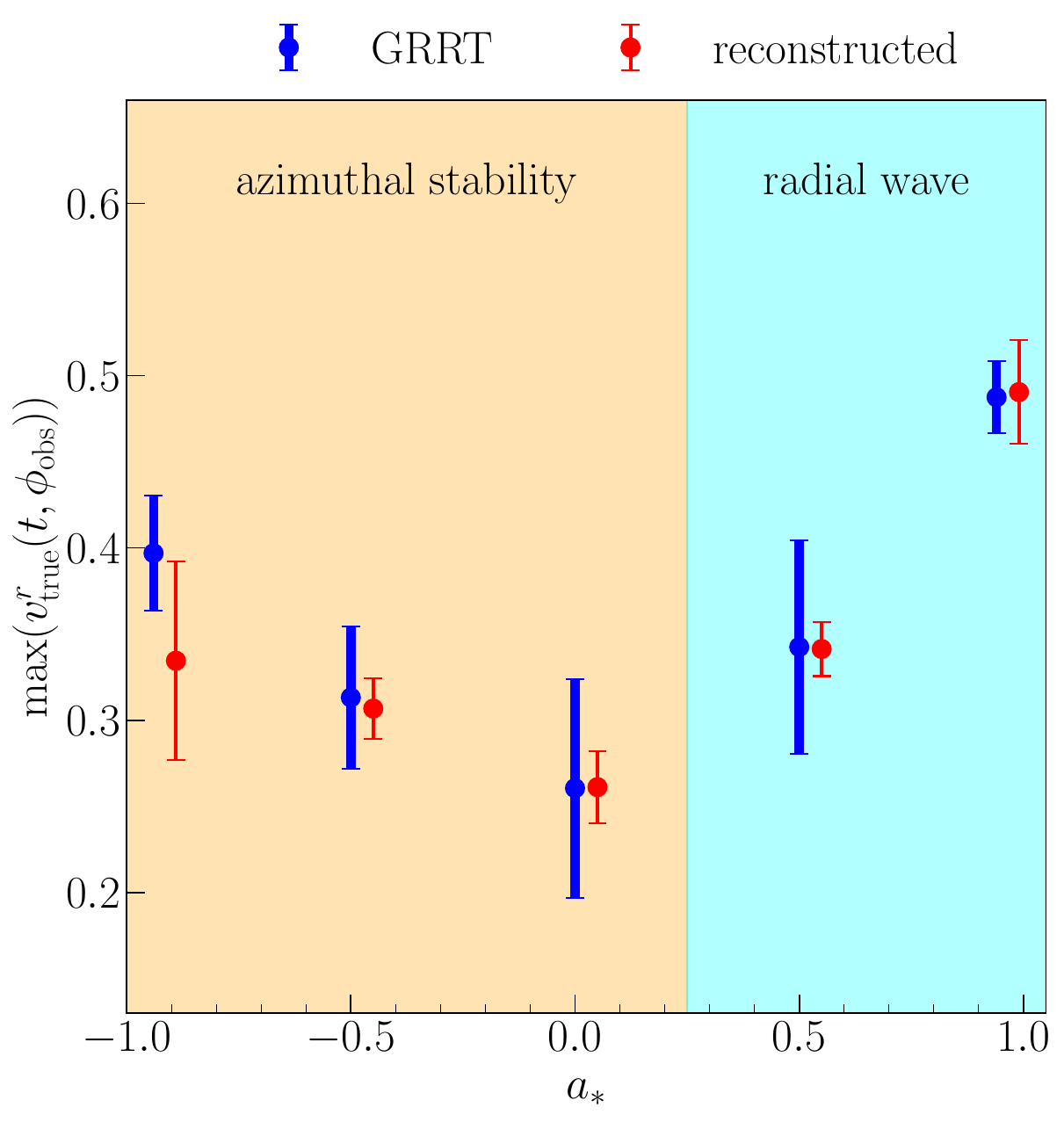}
\caption{
Spin dependency of the azimuthally maximum value of radial wave velocity $v^r_{\rm true}$ (refer to the bottom panels in Fig.~\ref{fig:phi_t_a_vr}), where the blue and red points represent the estimated values obtained from the GRRT and reconstructed images, respectively.
The top labels in the orange and cyan-colored regions indicate potential characteristics that can be used to constrain the spin within each spin range.
For clarity, the red points are slightly shifted to the right ($0.05$) in the horizontal direction.
}
\label{fig:summary_vr_spin}
\end{figure}

\begin{table}
\caption{
Summary of the spin dependency of the disk-jet dynamic. 
In this table, $\sigma_{\rm jet}(\phi_{\rm obs})$ and $\sigma_{\rm disk}(\phi_{\rm obs})$ denote the standard deviation of the peak azimuthal angles in the jet and the disk regions, respectively (refer to Fig.~\ref{fig:abh_phi_peak}).
Furthermore, $v^{r}_{\rm true}$ represents the maximum radial wave velocity along the azimuth (refer to Fig.~\ref{fig:summary_vr_spin}).
The values enclosed in parentheses are the estimates derived from the reconstructed images.
} \centering
\begin{tabular}{cccc}
\hline
$a_*$ & $\sigma_{\rm jet}(\phi_{\rm obs})~ [^\circ]$& $\sigma_{\rm disk}(\phi_{\rm obs})~ [^\circ]$ & $v^{r}_{\rm true}$ \\
\hline
0.9375 &  4  ( 4) &  13 ( 11) & 0.49$\pm$ 0.02 (0.49$\pm$ 0.03)\\
0.5 &  5  ( 6) &  16 ( 13) & 0.34$\pm$ 0.06 (0.34$\pm$ 0.02)\\
0.0 & 53  (52) &  53 ( 49) & 0.26$\pm$ 0.06 (0.26$\pm$ 0.02)\\
-0.5 & 42  (40) &  91 ( 89) & 0.31$\pm$ 0.04 (0.31$\pm$ 0.02)\\
-0.9375 & 72  (72) &  73 ( 73) & 0.40$\pm$ 0.03 (0.33$\pm$ 0.06)\\
\hline
\end{tabular}
\label{tab:summary_dynamics}
\end{table} 

By utilizing the spacetime diagram in the right panels of Fig.~\ref{fig:demo_spacetime_diagram}, we conducted an investigation into the temporal and radial development of the inhomogeneous waves in the jet region.
We estimated the time and azimuthal distribution of the number of waves $N$, relative brightness $f$, and radial velocity $v^r$ of each wave through the following steps:

\begin{enumerate}
\item 
Given an azimuthal angle $\phi$, we collected radial waves with $f(t,r_{\rm obs})>0.1$, that is, with an excess of $10\%$ in the relative temperature brightness.
Continuity of wave propagation is defined as a set of temporally and radially continuous points, spanning a time duration of more than $30~t_{\rm g}$.
Based on this process, we obtained a set of events $(t, r_{\rm obs})$ and the corresponding relative brightness $f$ of each wave train.
\item 
We repeated this procedure at each azimuthal angle, resulting in a set of waves $(t, r_{\rm obs}, \phi_{\rm obs}, f)_{j}$, where $i$ denotes the wave index $1\leq j\leq N_{\rm tot}$ and $N_{\rm tot}$ is the total number of detected waves.
%
\item 
We conducted a linear fitting to each wave $j$ using the least square method, such that the slope of the linear function $r_{\rm obs,~ fit}(t_{j})$ provides the radial velocity $v^r$.
Note that because the estimated velocity is projected to the image plane, the deprojected relativistic velocity $v^r_{\rm true}$ is calculated with the observed apparent one $v^r$ (e.g., \citealt{Falla2002}):
\begin{equation*}
v^r_{\rm true} = \frac{v^r}{\sin (180^\circ-i) + v^r\cos (180^\circ-i)}.
\end{equation*}
%
\item 
Finally, the velocity was averaged both temporally and azimuthally across sections with widths $\Delta t$ and $\Delta \phi$.
For the formers, we selected $\Delta t=200t_{\rm g}$, which corresponds to the propagation time in the radial jet region ($100~\uas\leq r_{\rm obs}<300~\uas$) when $v^r\geq 0.1$.
For the latter, we have chosen an azimuthal section, $\Delta \phi=30^\circ$, which encompasses the range spanned by azimuthal jet brightness for $a_*\geq 0.5$ (see Fig.\ref{fig:abh_phi_peak}), and is consistent with the width of the bottom histograms in Fig.\ref{fig:phi_t_Tb_jet}.
The validity of the radial velocity estimate was inspected using geometric jet models shown in Appendix~\ref{subsec:geometric}.
\end{enumerate}

Following steps (1)-(3), we investigated the radial wave properties in the jet region, and 
Fig.~\ref{fig:res_fit} shows the number of detected waves $N$ with respect to the relative brightness $f$.
The top and middle panels show the distributions of $N/{\rm max}(N)$ at the residual radius $r_{\rm obs}-r_{\rm obs,~fit}$ and the relative brightness of each wave, utilizing the GRRT (top panels) and reconstructed images (middle panels), respectively.
Clearly the maximum value of the relative brightness $f$ depends on the absolute value of the spin, and for $N/\max(N) = 0.1$ (white curves), $f$ extends to approximately $1.5, 1.4, 1.4, 1.3,$ and $1.6$ in the cases of $a_*=-0.9375,-0.5, 0.0, 0.5, $ and $0.9375$, respectively.
This behavior suggests the inhomogeneity of the radial waves increases with the magnitude of the black-hole spin.
The number of radial waves, $N_{\rm tot}$ (left value), monotonically increases with the absolute value of the spin in the case of the GRRT images (top panels), and a similar trend is seen also for the reconstructed images (middle panels).
However, they do not quite accurately capture the spin dependence observed in the GRRT images due to the uncertainties in the synthetic observations.
The bottom panels show the wave width of the fit lines.
Each component is radially symmetric and has a FWHM of $\sim 20~ \uas$, which is the restoring beam size of the image (see Section~\ref{subsec:imaging}).
The symmetric profiles, characterized by the FWHM, indicate the accurate detection of the center of jet waves using linear fitting. 
This also provides insight into the quality of the fitting process, with the validation detailed in Appendix~\ref{subsec:geometric}.

Figure~\ref{fig:phi_t_a_vr} presents the estimated radial velocities $v^r_{\rm true}$ for each spin.
The top and middle panels display the distributions of $v^r_{\rm true}$ at each azimuthal angle and time using the GRRT and reconstructed images, respectively.
The estimated velocities exhibit significant variations in both time and azimuthal directions for $a_*< 0.5$, while demonstrating more robust statistics for $a_*\geq 0.5$. 
This behavior is consistent with the spin dependence of the azimuthal variability of the jet discussed in Section~\ref{subsec:structure_variation}. 
The bottom panels show the median velocity with respect to the temporal direction for each azimuthal angle, where the error bars correspond to the median absolute deviations.
Note that the velocity peaks around the rotation axis of the black hole $(\phi=18^\circ)$ for all spins.

Finally, Fig.~\ref{fig:summary_vr_spin} summarizes many of the results presented so far and the relationship between spin and the maximum radial velocity in the azimuthal direction.
The estimated velocity, as indicated by the GRRT (blue) and reconstructed (red) images, demonstrates a monotonic dependency on the absolute spin value. 
These estimates allow us to distinguish between the cases of $a_*= 0.5$ and $a_*\geq 0.9375$. 
As discussed in Section~\ref{subsec:structure_variation}, the azimuthal variability of the jet and the disk enables differentiation in the case of $a_*\leq 0$ (the orange region in Fig.~\ref{fig:summary_vr_spin}). 
As a result, Fig.~\ref{fig:summary_vr_spin} provides evidence that the azimuthal and radial variations of the jet and the disk depend on the black-hole spin, thus enabling us to extract spacetime information (\autoref{tab:summary_dynamics}). 
Similar spin dependences are also observable at different inclination angles $i$, with the measurement of radial velocity and azimuthal variation providing constraints on $a_*$ and $i$ (Appendix~\ref{sec:inc}).




\section{Physical interpretation of jet radial waves}\label{sec:velocity_origin}
\begin{figure*}
\centering
\includegraphics[width=\linewidth]{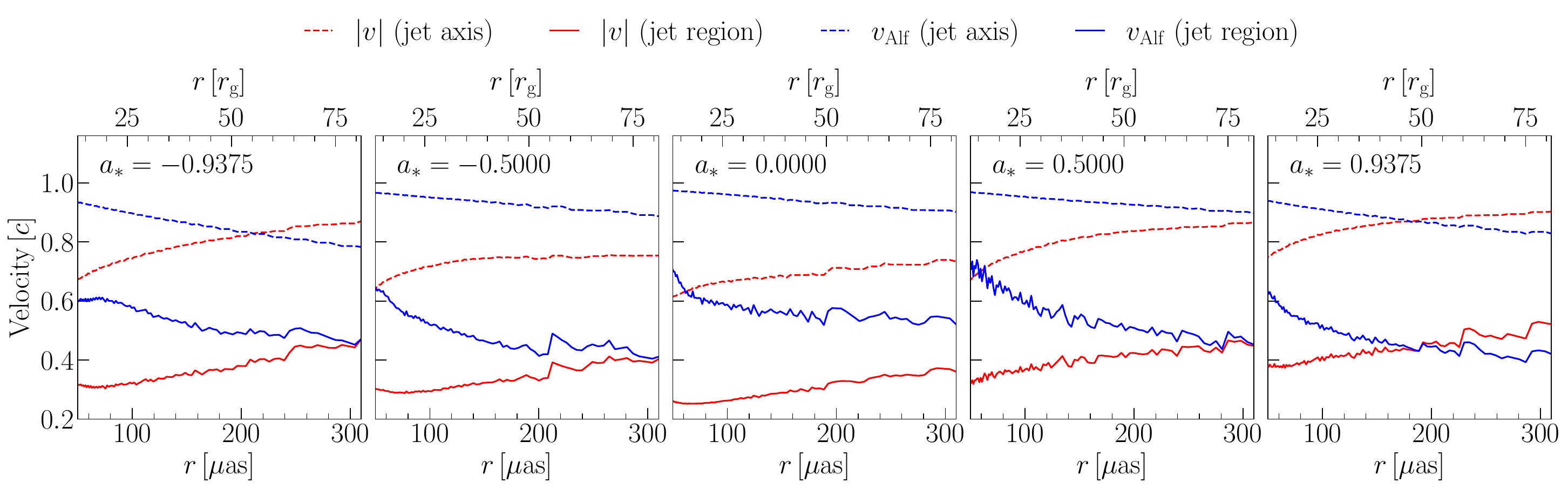}
\caption{
Radial distribution of the fluid velocity  ($|v|$, red curve) and Alfv{\'{e}}n velocity ($v_{\rm Alf}$, blue curve) averaged over time and around the jet for each spin value.
The dashed and solid curves correspond to the average around the vertical jet axis ($\sigma_{\rm B}>3$) and in the jet region (${\rm Be}>1.02$ and $\sigma_{B}<3$).
}
\label{fig:v_GRMHD}
\end{figure*}

\begin{figure}
\centering
\includegraphics[width=\linewidth]{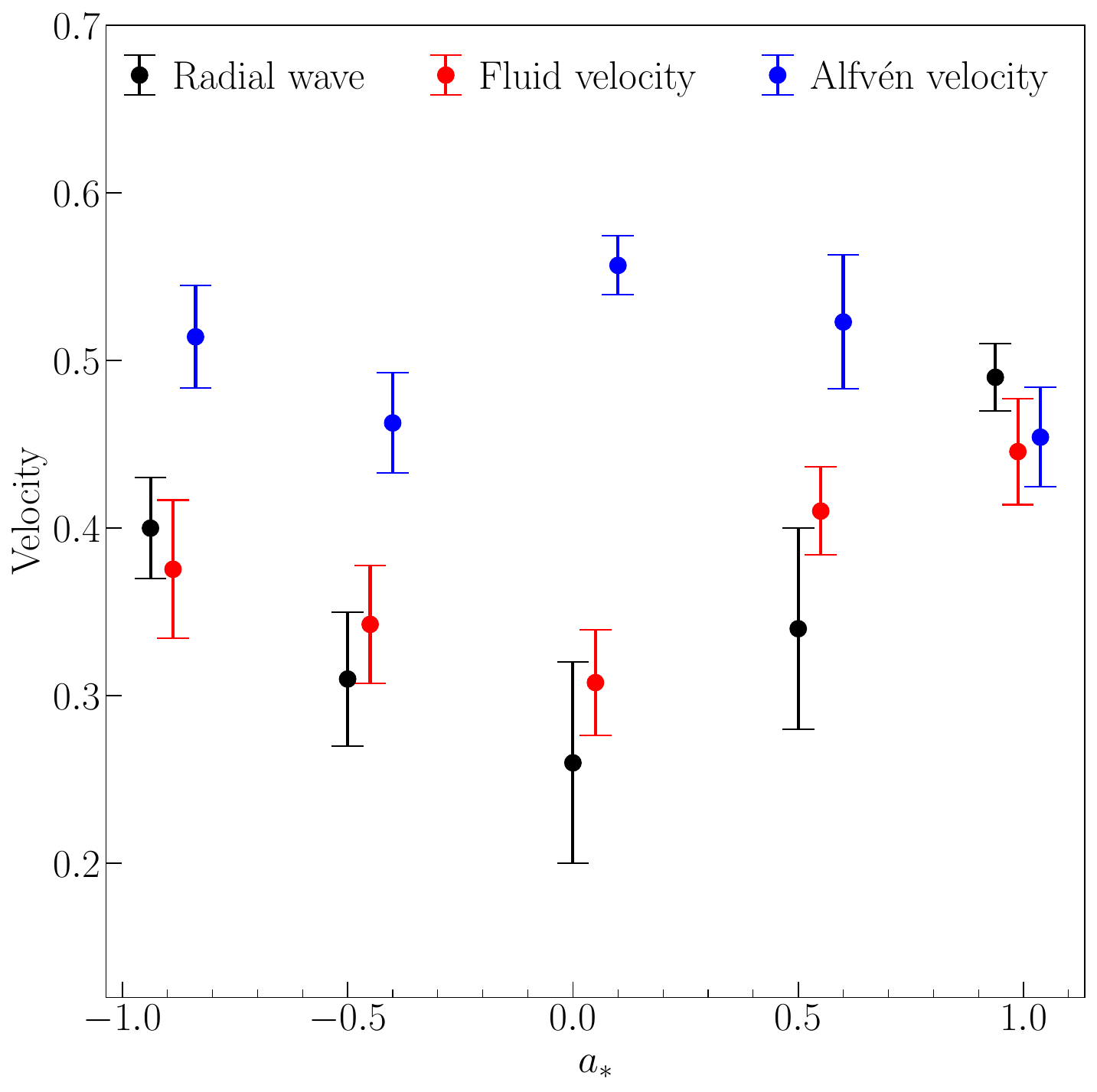}
\caption{
Spin dependency of the fluid and Alfv{\'{e}}n velocities is shown as averaged values in the jet region (represented by red and blue points, respectively), along with the radial wave velocity $v^r_{\rm true}$ (depicted in black, as detailed in Fig.~\ref{fig:summary_vr_spin}). For clarity, the positions of the red and blue points are slightly shifted to the right by increments of $0.05$ and $0.1$ in the horizontal direction, respectively.
}
\label{fig:vcomparison_GRMHD}
\end{figure}


The origin of radial jet waves can primarily be attributed to the motion
of relativistic plasma along the jet, a phenomenon both confined and
accelerated by the magnetic field. By examining the fluid velocities
($|v|=\sqrt{v_i v^i}$, where $v_i$ and $v^i$ are the covariant and
contravariant three velocities of the accretion flow) and Alfv{\'{e}}n
velocities ($v_{\rm Alf}=\sqrt{\sigma_{B}/(\sigma_{B}+h)}$), we
investigate the origin of the radial jet wave velocity derived in the
previous section. Here, $\sigma_{B}=b^2/\rho$ is the plasma
magnetization, $b$ is the norm of the magnetic field in the fluid frame,
and $\rho$ is the rest-mass density. Figure~\ref{fig:v_GRMHD} reports
radial distribution of the velocities averaged over time and around the
jet; the red and blue curves denote the fluid and Alfv{\'{e}}n
velocities, respectively. The dashed and solid lines indicate the
average values around the jet axis ($\sigma_B > 3$) and in the jet region
($\sigma_B < 3$ and $\rm Be > 1.02$), respectively. In each region and
each spin case, the fluid velocity (blue curves) exhibits a monotonic
increase with the radius due to the diminishing gravitational attraction.
Conversely, the Alfv{\'{e}}n velocity (red curves) decreases
monotonically with the radius as it moves farther from the center where
the magnetic field is most concentrated.

Both velocities have a monotonic dependency on the spin around the jet axis (see dashed curves). 
The fluid and Alfv{\'{e}}n velocity in the region averaged in the radial direction, are given by $|v| = 0.81, ~ 0.74,~ 0.70,~ 0.82,~ 0.87$ and $v_{\rm Alf} = 0.85,~ 0.93,~ 0.94,~ 0.93,~ 0.87~$, corresponding to the black-hole spins of $a_* = -0.9375,~ -0.5,~ 0.0,~ 0.5$, and $0.9375$, respectively.
These results indicate that the accretion flows around the jet region accelerated with the strong magnetic fields represented by the Alfv{\'{e}}n velocity ($v_{\rm Alf}\sim 0.90$) and the fluid velocity monotonically increases with the magnitude of the spin.

We further investigate the spin dependency of the fluid and Alfv{\'{e}}n velocities within the jet region.
In Fig.~\ref{fig:vcomparison_GRMHD}, we present a comparison between the spatially averaged fluid and Alfv{\'{e}}n velocities in the jet region (red and blue points) and the estimated velocity derived from the jet radial waves (Section~\ref{subsec:spin_vr}, depicted in black).
The accompanying error bars for the fluid and Alfv{\'{e}}n velocities represent the standard deviation. 
The fluid velocities align closely with the estimates from the jet radial waves. 
In contrast, the Alfv{\'{e}}n velocities do not exhibit a clear spin dependency.
It is further noted that the fluid velocity in the jet region is predominantly radial, with ratios of the radial three-velocity $\sqrt{v^r v_r}$ and $|v|$ averaged in the jet region are $85~\%,~ 90~\%,~ 93~\%,~ 92~\%,~ 95~\%$ for $a_* = -0.9375, -0.5, 0.0, 0.5$, and $0.9375$, respectively.
The results indicate that inhomogeneity within the jet serves as an effective marker, tracing the contribution of spin to the relativistic fluid motion.



\section{Summary and future prospects}\label{sec:summary}

We investigated the horizon-scale dynamics of the jet and the accretion disk and demonstrated the potential of black-hole spin measurement. 
We focused on the best-bet GRMHD models whose continuum spectra are consistent with that of the {\m87} observations within a small set of model parameters (\citealt{Fromm2022, Cruz2022}). 
The azimuthal variation of the jet and disk brightness reveals information about the spin direction, with the azimuthal angle of peak brightness showing temporal stability in cases of $a_*>0$ and instability when $a_*\leq 0$.
Furthermore, this variation has the potential to distinguish negative spin cases. 
In the jet region ($100~\uas\lesssim r_{\rm obs}\lesssim 300~ \uas$), inhomogeneous waves propagate radially with relativistic velocities ($0.26\lesssim v^r_{\rm true}\lesssim 0.50$). 
The radial wave velocity, as estimated from the GRRT images, shows a monotonic dependence on the absolute value of the black-hole spin. 
The reconstructed images from the ngEHT have demonstrated the detectability of these dynamics and the accuracy of the estimated black-hole spin.
This paper provides the essence of the spacetime measurement with the horizon-scale dynamics and future theoretical and observational projects.

A survey with a broader library of GRMHD simulations will provide a more comprehensive view of spin measurement under various possibilities of spacetime, magnetic field, accretion flow properties, and inclination angles.
For a survey with an expanded library of GRMHD simulations, it is essential to investigate each regime of parameter constraints with individual properties: continuum spectrum, visibility fitting, polarization, and jet-disk dynamics (\citealt{Fromm2022, EHT_M87_PaperVI, EHT_M87_PaperVII, Davelaar2023}). 
In addition to the expansion of model parameters, the dynamics survey has the potential to provide additional information on tilted disk parameters depending on the Lense-Thirring precession of the accretion disk and jet (e.g., \citealt{Liska2018, Chatterjee2020}).
%
%
We also focused on the simplest case without gain uncertainty and performed image reconstruction. 
To investigate the fiducial morphology under more realistic conditions, we need to conduct imaging surveys with numerous imaging parameters (\citealt{EHT_M87_PaperIV, EHT_SgrA_PaperIII}) with the more realistic data including gain uncertainties, polarization leakages, and instrumental effects.

Alongside the prospective utility of ngEHT observations, it is worth noting that the annual azimuthal variation in the photon ring, as observed through current and impending EHT studies, is also synergetic with our proposed method.
The annual EHT observations, which offer less dense $(u,v)$ coverages compared to ngEHT, are either under analysis or in the planning stages to investigate variations of the photon ring.
Particularly, RML imaging, as demonstrated with the 2017 EHT data, has furnished an azimuthal angle assessment of the photon ring, achieving an accuracy of approximately $\sim 20^\circ$ (\citealt{EHT_M87_PaperIV}).
By continuous observations of the azimuthal angle variabilities and capturing the temporal evolution of images within the disk region, discerning differences in the azimuthal angle variation correlated to the black hole's spin direction (roughly $60^\circ$) will become attainable.
Conversely, the task of differentiating small differences in azimuthal angle variabilities between ($a_*=0.5$) and ($a_*=0.9375$) (around $2-3^\circ$, as outlined in \autoref{tab:summary_dynamics}) is challenging. 
The observation of the jet's temporal progression, augmented by the denser \((u,v)\) coverage facilitated by ngEHT observations, will be an essential requirement for constraining the magnitudes of the spin.

The quality of horizon-scale images has improved significantly due to
recent advancements in imaging techniques (e.g., \citealt{Chael2022,
  Muller2022, Arras2022}). These developments potentially pave the way
for the application of our spin measurement methodology to EHT
observations from 2021 onwards. These results, in conjunction with EHT
observations, are expected to synergize with current and future
multi-wavelength observations (\citealt{Lu2023}). By utilizing our
methodology in future EHT observations, theoretical and observational
studies will be able to achieve a much more comprehensive measurement of
the black-hole spacetime of {\m87}. \\

\textit{Acknowledgements.} We thank Prashant Kocherlakota and Indu Kalpa
Dihingia for their insightful comments on this research. This research
is supported by the European Research Council for the Advanced Grant
``JETSET: Launching, propagation and emission of relativistic jets from
binary mergers and across mass scales'' (Grant No. 884631). CMF is
supported by the DFG research grant ``Jet physics on horizon scales and
beyond" (Grant No. FR 4069/2-1). YM is supported by the National
Natural Science Foundation of China (Grant No. 12273022), the Shanghai
Municipality orientation program of basic research for international scientists 
(Grant No. 22JC1410600), and the National Key R\&D Program of China 
(Grant No. 2023YFE0101200).
LR acknowledges the Walter Greiner Gesellschaft zur
F\"orderung der physikalischen Grundlagenforschung e.V. through the Carl
W. Fueck Laureatus Chair. The simulations were performed on GOETHE-HLR
LOEWE at the CSC-Frankfurt, Calea and Iboga at ITP Frankfurt and Pi2.0
and Siyuan Mark-I in Shanghai Jiao Tong University. Software: {\bhac}
(https: //bhac.science/) (\citealt{Porth2017}), {\bhoss}
(\citealt{Younsi2020}), {\ehtim} (https://achael.github.io/eht-imaging/)
(\citealt{Chael2018_Imaging}), and {\smili}
(https://github.com/astrosmili/smili) (\citealt{Akiyama2017a,
  Akiyama2017b}).


\newpage

\begin{appendix}

\section{Inspection of radial velocity estimates with geometrical jet models}\label{subsec:geometric}

\begin{figure}
\centering
\includegraphics[width=\linewidth]{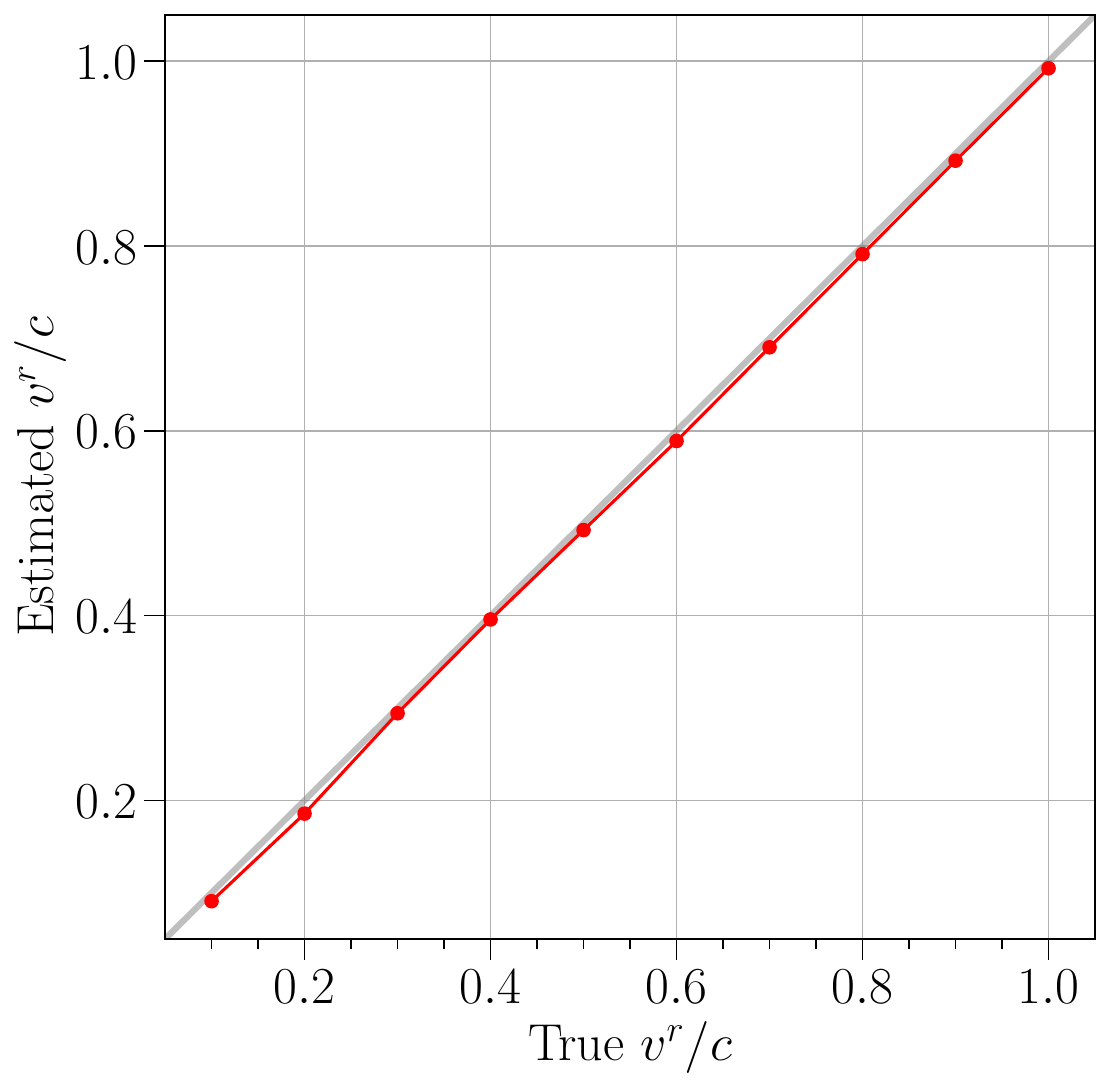}
\caption{
The radial velocity of the geometric jets as defined in Section~\ref{subsec:geometric}. 
The gray curve represents the input (or 'true') radial velocity, while the red curve represents the estimated radial velocity. 
}
\label{fig:geometric}
\end{figure}

In Section~\ref{subsec:spin_vr}, we introduced a method for estimating the radial wave velocity of jet inhomogeneity. 
To validate these velocity estimates, we applied the method to simple geometric jet models and evaluated the accuracy of the resulting velocity estimates. 
The intensity distribution of the geometric model is given by
\begin{eqnarray*}
    I(t, r,\phi) &=& I_{0}(t)G(t,r,\phi)\exp\left[-\left(\frac{r-v^r t}{\Delta r}\right)^2\right]~ (|\phi|<\Delta \phi), 
\end{eqnarray*}
where $v^r$ is the input velocity, $\Delta r$ and $\Delta \phi$ represent the radial and azimuthal width of the geometric wave, and $G(t,r,\phi)$ is a random Gaussian distribution with a mean and standard deviation of one. 
The intensity $I_0(t)$ is defined such that the total flux equals $1.0~{\rm Jy}$ at each time.

Figure~\ref{fig:geometric} compares the estimated velocities to the input
(true) velocities, showcasing typical examples with $\Delta r=20~\uas$
and $\Delta\phi =40^\circ$. The velocity estimates are derived from the
input images following the same strategy described in
Section~\ref{subsec:spin_vr}. Across the input velocity space in this
paper ($0.2\leq v^r \leq 1$), the estimated velocities accurately
reproduce the ground-truth values with an error of about $\Delta v^r\sim
0.01$.


\section{Inclination angle dependency}\label{sec:inc}
\begin{figure}
\centering
\includegraphics[width=\linewidth]{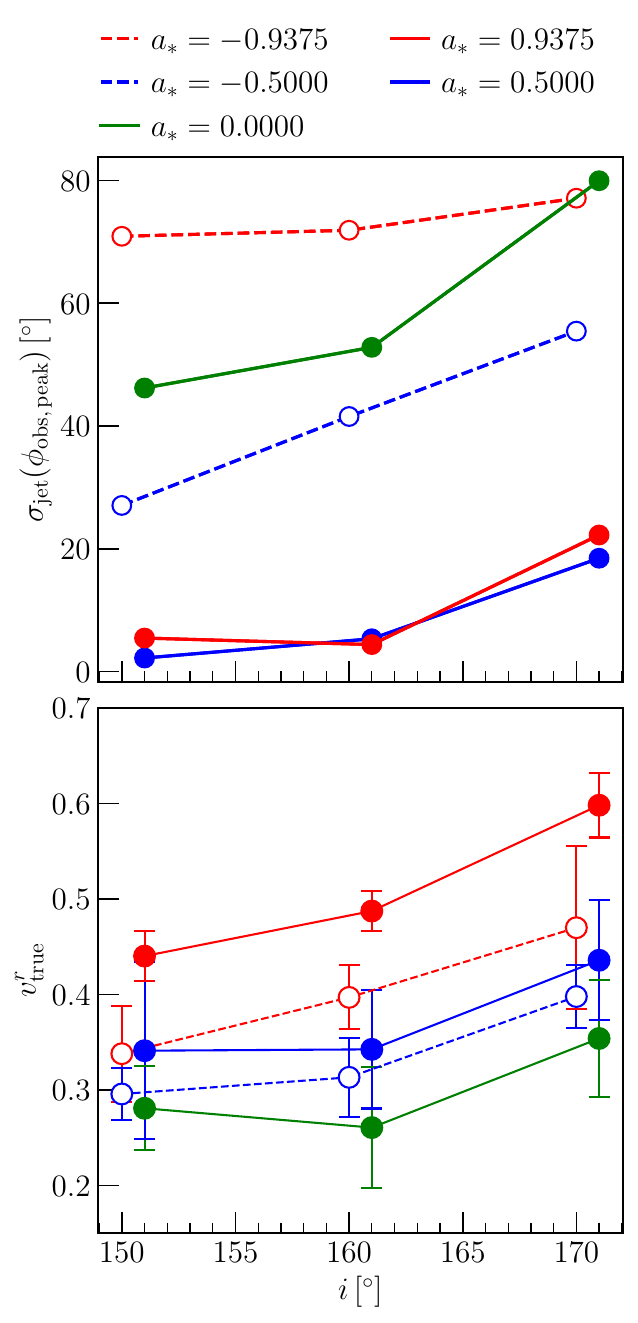}
\caption{
The inclination angle dependence of the estimated azimuthal variation (top panel) and radial velocity (bottom panel).
For clarity, the curves in the case of $a_*\geq 0$ are slightly ($1^\circ$) shifted in the horizontal direction.
}
\label{fig:inc_jet}
\end{figure}

The main text focuses on the best-bet model parameters, which reproduce the continuum spectrum of {\m87} at a fixed viewing angle ($i=160^\circ$).
In this appendix, we demonstrate that the broad spin dependency of both the azimuthal variability of the jet (Section~\ref{subsec:structure_variation}) $\sigma (\phi_{\rm obs,~ peak})$ and radial wave velocity $v^r_{\rm true}$ (Section~\ref{subsec:spin_vr}) is observable at different inclination angles.

In Fig.~\ref{fig:inc_jet}, we show the inclination angle dependency of
the azimuthal variation $\sigma (\phi_{\rm obs,~ peak})$ and the radial
wave velocity $v^r_{\rm true}$ for each spin (the red, blue, and green
curves represent the cases of $|a_*|=0.9375, 0.5,$ and $0$, respectively)
with the GRRT images. The transition value of $\sigma (\phi_{\rm obs,~
  peak})$ between $a_*\leq 0$ and $a_*>0$ is $\sigma(\phi_{\rm obs,~
  peak})\sim 30^\circ$ in the case of $150^\circ\leq i\leq 170^\circ$.
The radial velocity monotonically increases with $|a_*|$, indicating that
the inhomogeneous jet profile propagates more rapidly with the absolute
value of the spin. These two features are consistent across all
inclination angles. Thus, the measurement of the radial velocity and
azimuthal variability could potentially constrain both the spin value and
the viewing angle.


\end{appendix}

\bibliographystyle{yahapj}
\bibliography{aeireferences}

\begin{thebibliography}{}
\providecommand\natexlab[1]{#1}
\providecommand\JournalTitle[1]{#1}

\bibitem[{{Abramowicz} {et~al.}(2003){Abramowicz}, {Karas}, {Kluzniak}, {Lee},
  \& {Rebusco}}]{Abramowicz2003}
{Abramowicz}, M.~A., {Karas}, V., {Kluzniak}, W., {Lee}, W.~H., \& {Rebusco},
  P. 2003,
  \href{http://adsabs.harvard.edu/abs/2003PASJ...55..467A}{\JournalTitle{\pasj},
  55, 467}

\bibitem[{{Akiyama} {et~al.}(2017{\natexlab{a}}){Akiyama}, {Kuramochi},
  {Ikeda}, {Fish}, {Tazaki}, {Honma}, {Doeleman}, {Broderick}, {Dexter},
  {Mo{\'s}cibrodzka}, {Bouman}, {Chael}, \& {Zaizen}}]{Akiyama2017a}
{Akiyama}, K., {Kuramochi}, K., {Ikeda}, S., {et~al.} 2017{\natexlab{a}},
  \href{http://dx.doi.org/10.3847/1538-4357/aa6305}{\JournalTitle{\apj}, 838,
  1}

\bibitem[{{Akiyama} {et~al.}(2017{\natexlab{b}}){Akiyama}, {Ikeda}, {Pleau},
  {Fish}, {Tazaki}, {Kuramochi}, {Broderick}, {Dexter}, {Mo{\'s}cibrodzka},
  {Gowanlock}, {Honma}, \& {Doeleman}}]{Akiyama2017b}
{Akiyama}, K., {Ikeda}, S., {Pleau}, M., {et~al.} 2017{\natexlab{b}},
  \href{http://dx.doi.org/10.3847/1538-3881/aa6302}{\JournalTitle{\aj}, 153,
  159}

\bibitem[{Arras {et~al.}(2022)Arras, Frank, Haim, Knollmüller,
  {et~al.}}]{Arras2022}
Arras, P., Frank, P., Haim, P., Knollmüller, J., {et~al.} 2022,
  \href{https://ui.adsabs.harvard.edu/abs/2022NatAs...6..259A/abstract}{\JournalTitle{Nature
  Astronomy}, 259}

\bibitem[{{Blandford} \& {Znajek}(1977)}]{Blandford1977}
{Blandford}, R.~D., \& {Znajek}, R.~L. 1977, \JournalTitle{\mnras}, 179, 433

\bibitem[{{Chael} {et~al.}(2022){Chael}, {Issaoun}, {Pesce}, {Johnson},
  {et~al.}}]{Chael2022}
{Chael}, A., {Issaoun}, S., {Pesce}, D.~W., {Johnson}, M.~D., {et~al.} 2022,
  \href{https://ui.adsabs.harvard.edu/abs/2022arXiv221012226C/abstract}{\JournalTitle{Galaxies},
  11, 23}

\bibitem[{{Chael} {et~al.}(2021){Chael}, {Johnson}, \& {Lupsasca}}]{Chael2021}
{Chael}, A., {Johnson}, M.~D., \& {Lupsasca}, A. 2021,
  \href{http://dx.doi.org/10.3847/1538-4357/ac09ee}{\JournalTitle{\apj}, 918,
  6}

\bibitem[{{Chael} {et~al.}(2018{\natexlab{a}}){Chael}, {Rowan}, {Narayan},
  {Johnson}, \& {Sironi}}]{Chael2018}
{Chael}, A., {Rowan}, M., {Narayan}, R., {Johnson}, M., \& {Sironi}, L.
  2018{\natexlab{a}},
  \href{http://dx.doi.org/10.1093/mnras/sty1261}{\JournalTitle{\mnras}, 478,
  5209}

\bibitem[{{Chael} {et~al.}(2018{\natexlab{b}}){Chael}, {Johnson}, {Bouman},
  {Blackburn}, {Akiyama}, \& {Narayan}}]{Chael2018_Imaging}
{Chael}, A.~A., {Johnson}, M.~D., {Bouman}, K.~L., {et~al.} 2018{\natexlab{b}},
  \href{http://dx.doi.org/10.3847/1538-4357/aab6a8}{\JournalTitle{\apj}, 857,
  23}

\bibitem[{{Chatterjee} {et~al.}(2020){Chatterjee}, {Younsi}, {Liska},
  {Tchekhovskoy}, {Markoff}, {Yoon}, {van Eijnatten}, {Hesp}, {Ingram}, \& {van
  der Klis}}]{Chatterjee2020}
{Chatterjee}, K., {Younsi}, Z., {Liska}, M., {et~al.} 2020,
  \href{http://dx.doi.org/10.1093/mnras/staa2718}{\JournalTitle{\mnras}, 499,
  362}

\bibitem[{Cruz-Osorio {et~al.}(2021)Cruz-Osorio, Gimeno-Soler, Font,
  De~Laurentis, \& Mendoza}]{Cruz2021}
Cruz-Osorio, A., Gimeno-Soler, S., Font, J.~A., De~Laurentis, M., \& Mendoza,
  S. 2021,
  \href{http://dx.doi.org/10.1103/PhysRevD.103.124009}{\JournalTitle{Phys. Rev.
  D}, 103, 124009}

\bibitem[{{Cruz-Osorio} {et~al.}(2022){Cruz-Osorio}, {Fromm}, {Mizuno},
  {Nathanail}, {Younsi}, {Porth}, {Davelaar}, {Falcke}, {Kramer}, \&
  {Rezzolla}}]{Cruz2022}
{Cruz-Osorio}, A., {Fromm}, C.~M., {Mizuno}, Y., {et~al.} 2022,
  \href{http://dx.doi.org/10.1038/s41550-021-01506-w}{\JournalTitle{Nature
  Astronomy}, 6, 103}

\bibitem[{{Curtis}(1918)}]{Curtis1918}
{Curtis}, H.~D. 1918, \JournalTitle{Publications of Lick Observatory}, 13, 9

\bibitem[{{Davelaar} {et~al.}(2019){Davelaar}, {Olivares}, {Porth},
  {Bronzwaer}, {Janssen}, {Roelofs}, {Mizuno}, {Fromm}, {Falcke}, \&
  {Rezzolla}}]{Davelaar2019}
{Davelaar}, J., {Olivares}, H., {Porth}, O., {et~al.} 2019,
  \href{http://dx.doi.org/10.1051/0004-6361/201936150}{\JournalTitle{\aap},
  632, A2}

\bibitem[{{Davelaar} {et~al.}(2023){Davelaar}, {Ripperda}, {Sironi},
  {Philippov}, {Olivares}, {Porth}, {van den Berg}, {Bronzwaer}, {Chatterjee},
  \& {Liska}}]{Davelaar2023}
{Davelaar}, J., {Ripperda}, B., {Sironi}, L., {et~al.} 2023,
  \href{https://ui.adsabs.harvard.edu/abs/2023arXiv230907963D/abstract}{\JournalTitle{https://arxiv.org/abs/2309.07963}}

\bibitem[{{Doeleman} {et~al.}(2019){Doeleman}, {Blackburn}, {Dexter}, {Gomez},
  {Johnson}, {Palumbo}, {Weintroub}, {Farah}, {Fish}, {Loinard}, {Lonsdale},
  {Narayanan}, {Patel}, {Pesce}, {Raymond}, {Tilanus}, {Wielgus}, {Akiyama},
  {Bower}, {Broderick}, {Deane}, {Fromm}, {Gammie}, {Gold}, {Janssen},
  {Kawashima}, {Krichbaum}, {Marrone}, {Matthews}, {Mizuno}, {Rezzolla},
  {Roelofs}, {Ros}, {Savolainen}, {Yuan}, {Zhao}, {Blackburn}, {Doeleman},
  {Dexter}, {Gomez}, {Johnson}, {Palumbo}, {Weintroub}, {Farah}, {Fish},
  {Loinard}, {Lonsdale}, {Narayanan}, {Patel}, {Pesce}, {Raymond}, {Tilanus},
  {Wielgus}, {Akiyama}, {Bower}, {Broderick}, {Deane}, {Fromm}, {Gammie},
  {Gold}, {Janssen}, {Kawashima}, {Krichbaum}, {Marrone}, {Matthews}, {Mizuno},
  {Rezzolla}, {Roelofs}, {Ros}, {Savolainen}, {Yuan}, \& {Zhao}}]{Doeleman2019}
{Doeleman}, S., {Blackburn}, L., {Dexter}, J., {et~al.} 2019,
  \JournalTitle{BAAS}, 51, 256

\bibitem[{{Doeleman} {et~al.}(2012){Doeleman}, {Fish}, {Schenck}, {Beaudoin},
  {Blundell}, {Bower}, {Broderick}, {Chamberlin}, {Freund}, {Friberg},
  {Gurwell}, {Ho}, {Honma}, {Inoue}, {Krichbaum}, \& et~al.}]{Doeleman2012}
{Doeleman}, S.~S., {Fish}, V.~L., {Schenck}, D.~E., {et~al.} 2012,
  \href{http://dx.doi.org/10.1126/science.1224768}{\JournalTitle{Science}, 338,
  355}

\bibitem[{{Event Horizon Telescope Collaboration}
  {et~al.}(2019{\natexlab{a}}){Event Horizon Telescope Collaboration},
  {Akiyama}, {Alberdi}, {Alef}, {Asada}, {Azulay}, {Baczko}, {Ball},
  {Balokovi{\'c}}, {Barrett}, {et~al.}}]{EHT_M87_PaperI}
{Event Horizon Telescope Collaboration}, {Akiyama}, K., {Alberdi}, A., {et~al.}
  2019{\natexlab{a}},
  \href{http://dx.doi.org/10.3847/2041-8213/ab0ec7}{\JournalTitle{\apjl}, 875,
  L1}

\bibitem[{{Event Horizon Telescope Collaboration}
  {et~al.}(2019{\natexlab{b}}){Event Horizon Telescope Collaboration},
  {Akiyama}, {Alberdi}, {Alef}, {Asada}, {Azulay}, {Baczko}, {Ball},
  {Balokovi{\'c}}, {Barrett}, {et~al.}}]{EHT_M87_PaperII}
---. 2019{\natexlab{b}},
  \href{http://dx.doi.org/10.3847/2041-8213/ab0c96}{\JournalTitle{\apjl}, 875,
  L2}

\bibitem[{{Event Horizon Telescope Collaboration}
  {et~al.}(2019{\natexlab{c}}){Event Horizon Telescope Collaboration},
  {Akiyama}, {Alberdi}, {Alef}, {Asada}, {Azulay}, {Baczko}, {Ball},
  {Balokovi{\'c}}, {Barrett}, {et~al.}}]{EHT_M87_PaperIII}
---. 2019{\natexlab{c}},
  \href{http://dx.doi.org/10.3847/2041-8213/ab0c57}{\JournalTitle{\apjl}, 875,
  L3}

\bibitem[{{Event Horizon Telescope Collaboration}
  {et~al.}(2019{\natexlab{d}}){Event Horizon Telescope Collaboration},
  {Akiyama}, {Alberdi}, {Alef}, {Asada}, {Azulay}, {Baczko}, {Ball},
  {Balokovi{\'c}}, {Barrett}, {et~al.}}]{EHT_M87_PaperIV}
---. 2019{\natexlab{d}},
  \href{http://dx.doi.org/10.3847/2041-8213/ab0e85}{\JournalTitle{\apjl}, 875,
  L4}

\bibitem[{{Event Horizon Telescope Collaboration}
  {et~al.}(2019{\natexlab{e}}){Event Horizon Telescope Collaboration},
  {Akiyama}, {Alberdi}, {Alef}, {Asada}, {Azulay}, {Baczko}, {Ball},
  {Balokovi{\'c}}, {Barrett}, {et~al.}}]{EHT_M87_PaperV}
---. 2019{\natexlab{e}},
  \href{http://dx.doi.org/10.3847/2041-8213/ab0f43}{\JournalTitle{\apjl}, 875,
  L5}

\bibitem[{{Event Horizon Telescope Collaboration}
  {et~al.}(2019{\natexlab{f}}){Event Horizon Telescope Collaboration},
  {Akiyama}, {Alberdi}, {Alef}, {Asada}, {Azulay}, {Baczko}, {Ball},
  {Balokovi{\'c}}, {Barrett}, {et~al.}}]{EHT_M87_PaperVI}
---. 2019{\natexlab{f}},
  \href{http://dx.doi.org/10.3847/2041-8213/ab1141}{\JournalTitle{\apjl}, 875,
  L6}

\bibitem[{{Event Horizon Telescope Collaboration}
  {et~al.}(2021{\natexlab{a}}){Event Horizon Telescope Collaboration},
  {Akiyama}, {Algaba}, {Alberdi}, {Alef}, {Anantua}, {Asada}, {Azulay},
  {Baczko}, {Ball}, {Balokovi{\'c}}, {Barrett}, {et~al.}}]{EHT_M87_PaperVII}
{Event Horizon Telescope Collaboration}, {Akiyama}, K., {Algaba}, J.~C.,
  {et~al.} 2021{\natexlab{a}},
  \href{http://dx.doi.org/10.3847/2041-8213/abe71d}{\JournalTitle{\apjl}, 910,
  L12}

\bibitem[{{Event Horizon Telescope Collaboration}
  {et~al.}(2021{\natexlab{b}}){Event Horizon Telescope Collaboration},
  {Akiyama}, {Algaba}, {Alberdi}, {Alef}, {Anantua}, {Asada}, {Azulay},
  {Baczko}, {Ball}, {Balokovi{\'c}}, {Barrett}, {et~al.}}]{EHT_M87_PaperVIII}
---. 2021{\natexlab{b}},
  \href{http://dx.doi.org/10.3847/2041-8213/abe4de}{\JournalTitle{\apjl}, 910,
  L13}

\bibitem[{{Event Horizon Telescope Collaboration}
  {et~al.}(2022{\natexlab{a}}){Event Horizon Telescope Collaboration},
  {Akiyama}, {Alberdi}, {Alef}, {Algaba}, {Anantua}, {Asada}, {Azulay}, {Bach},
  {Baczko}, {Ball}, {Balokovi{\'c}}, {Barrett}, {Baub{\"o}ck}, {Benson},
  {Bintley}, {Blackburn}, {Blundell}, {Bouman}, {Bower}, {Boyce}, {Bremer},
  {Brinkerink}, {Brissenden}, {Britzen}, {Broderick}, {Broguiere}, {Bronzwaer},
  {Bustamante}, {Byun}, {Carlstrom}, {Ceccobello}, {Chael}, {Chan},
  {Chatterjee}, {Chatterjee}, {Chen}, {Chen}, {Cheng}, {Cho}, {Christian},
  {Conroy}, {Conway}, {Cordes}, {Crawford}, {Crew}, {Cruz-Osorio},
  {et~al.}}]{EHT_SgrA_PaperI}
{Event Horizon Telescope Collaboration}, {Akiyama}, K., {Alberdi}, A., {et~al.}
  2022{\natexlab{a}},
  \href{http://dx.doi.org/10.3847/2041-8213/ac6674}{\JournalTitle{\apjl}, 930,
  L12}

\bibitem[{{Event Horizon Telescope Collaboration}
  {et~al.}(2022{\natexlab{b}}){Event Horizon Telescope Collaboration},
  {Akiyama}, {Alberdi}, {Alef}, {Carlos Algaba}, {Anantua}, {Asada}, {Azulay},
  {Bach}, {Baczko}, {Ball}, {Balokovi{\'c}}, {Barrett}, {Baub{\"o}ck},
  {Benson}, {Bintley}, {Blackburn}, {Blundell}, {Bouman}, {Bower}, {Boyce},
  {Bremer}, {Brinkerink}, {Brissenden}, {Britzen}, {Broderick}, {Broguiere},
  {Bronzwaer}, {Bustamante}, {Byun}, {Carlstrom}, {Ceccobello}, {Chael},
  {Chan}, {Chatterjee}, {Chatterjee}, {Chen}, {Chen}, {Cheng}, {Cho},
  {Christian}, {Conroy}, {Conway}, {Cordes}, {Crawford}, {Crew}, {Cruz-Osorio},
  {et~al.}}]{EHT_SgrA_PaperII}
---. 2022{\natexlab{b}},
  \href{http://dx.doi.org/10.3847/2041-8213/ac6675}{\JournalTitle{\apjl}, 930,
  L13}

\bibitem[{{Event Horizon Telescope Collaboration}
  {et~al.}(2022{\natexlab{c}}){Event Horizon Telescope Collaboration},
  {Akiyama}, {Alberdi}, {Alef}, {Carlos Algaba}, {Anantua}, {Asada}, {Azulay},
  {Bach}, {Baczko}, {Ball}, {Balokovi{\'c}}, {Barrett}, {Baub{\"o}ck},
  {Benson}, {Bintley}, {Blackburn}, {Blundell}, {Bouman}, {Bower}, {Boyce},
  {Bremer}, {Brinkerink}, {Brissenden}, {Britzen}, {Broderick}, {Broguiere},
  {Bronzwaer}, {Bustamante}, {Byun}, {Carlstrom}, {Ceccobello}, {Chael},
  {Chan}, {Chatterjee}, {Chatterjee}, {Chen}, {Chen}, {Cheng}, {Cho},
  {Christian}, {Conroy}, {Conway}, {Cordes}, {Crawford}, {Crew}, {Cruz-Osorio},
  {et~al.}}]{EHT_SgrA_PaperIII}
---. 2022{\natexlab{c}},
  \href{http://dx.doi.org/10.3847/2041-8213/ac6429}{\JournalTitle{\apjl}, 930,
  L14}

\bibitem[{{Event Horizon Telescope Collaboration}
  {et~al.}(2022{\natexlab{d}}){Event Horizon Telescope Collaboration},
  {Akiyama}, {Alberdi}, {Alef}, {Carlos Algaba}, {Anantua}, {Asada}, {Azulay},
  {Bach}, {Baczko}, {Ball}, {Balokovi{\'c}}, {Barrett}, {Baub{\"o}ck},
  {Benson}, {Bintley}, {Blackburn}, {Blundell}, {Bouman}, {Bower}, {Boyce},
  {Bremer}, {Brinkerink}, {Brissenden}, {Britzen}, {Broderick}, {Broguiere},
  {Bronzwaer}, {Bustamante}, {Byun}, {Carlstrom}, {Ceccobello}, {Chael},
  {Chan}, {Chatterjee}, {Chatterjee}, {Chen}, {Chen}, {Cheng}, {Cho},
  {Christian}, {Conroy}, {Conway}, {Cordes}, {Crawford}, {Crew}, {Cruz-Osorio},
  {et~al.}}]{EHT_SgrA_PaperIV}
---. 2022{\natexlab{d}},
  \href{http://dx.doi.org/10.3847/2041-8213/ac6736}{\JournalTitle{\apjl}, 930,
  L15}

\bibitem[{{Event Horizon Telescope Collaboration}
  {et~al.}(2022{\natexlab{e}}){Event Horizon Telescope Collaboration},
  {Akiyama}, {Alberdi}, {Alef}, {Carlos Algaba}, {Anantua}, {Asada}, {Azulay},
  {Bach}, {Baczko}, {Ball}, {Balokovi{\'c}}, {Barrett}, {Baub{\"o}ck},
  {Benson}, {Bintley}, {Blackburn}, {Blundell}, {Bouman}, {Bower}, {Boyce},
  {Bremer}, {Brinkerink}, {Brissenden}, {Britzen}, {Broderick}, {Broguiere},
  {Bronzwaer}, {Bustamante}, {Byun}, {Carlstrom}, {Ceccobello}, {Chael},
  {Chan}, {Chatterjee}, {Chatterjee}, {Chen}, {Chen}, {Cheng}, {Cho},
  {Christian}, {Conroy}, {Conway}, {Cordes}, {Crawford}, {Crew}, {Cruz-Osorio},
  {et~al.}}]{EHT_SgrA_PaperV}
---. 2022{\natexlab{e}},
  \href{http://dx.doi.org/10.3847/2041-8213/ac6672}{\JournalTitle{\apjl}, 930,
  L16}

\bibitem[{{Event Horizon Telescope Collaboration}
  {et~al.}(2022{\natexlab{f}}){Event Horizon Telescope Collaboration},
  {Akiyama}, {Alberdi}, {Alef}, {Algaba}, {Anantua}, {Asada}, {Azulay}, {Bach},
  {Baczko}, {Ball}, {Balokovi{\'c}}, {Barrett}, {Baub{\"o}ck}, {Benson},
  {Bintley}, {Blackburn}, {Blundell}, {Bouman}, {Bower}, {Boyce}, {Bremer},
  {Brinkerink}, {Brissenden}, {Britzen}, {Broderick}, {Broguiere}, {Bronzwaer},
  {Bustamante}, {Byun}, {Carlstrom}, {Ceccobello}, {Chael}, {Chan},
  {Chatterjee}, {Chatterjee}, {Chen}, {Chen}, {Cheng}, {Cho}, {Christian},
  {Conroy}, {Conway}, {Cordes}, {Crawford}, {Crew}, {Cruz-Osorio}, {Cui},
  {Davelaar}, {De Laurentis}, {Deane}, {Dempsey}, {Desvignes}, {Dexter},
  {Dhruv}, {Doeleman}, {Dougal}, {Dzib}, {Eatough}, {Emami}, {Falcke}, {Farah},
  {Fish}, {Fomalont}, {Ford}, {Fraga-Encinas}, {Freeman}, {Friberg}, {Fromm},
  {Fuentes}, {Galison}, {Gammie}, {Garc{\'\i}a}, {Gentaz}, {Georgiev}, {Goddi},
  {Gold}, {G{\'o}mez-Ruiz}, {G{\'o}mez}, {Gu}, {Gurwell}, {Hada}, {Haggard},
  {Haworth}, {Hecht}, {Hesper}, {Heumann}, {Ho}, {Ho}, {Honma}, {Huang},
  {Huang}, {Hughes}, {Ikeda}, {Impellizzeri}, {Inoue}, {Issaoun}, {James},
  {Jannuzi}, {Janssen}, {Jeter}, {Jiang}, {Jim{\'e}nez-Rosales}, {Johnson},
  {Jorstad}, {Joshi}, {Jung}, {Karami}, {Karuppusamy}, {Kawashima}, {Keating},
  {Kettenis}, {Kim}, {Kim}, {Kim}, {Kim}, {Kino}, {Koay}, {Kocherlakota},
  {Kofuji}, {Koch}, {Koyama}, {Kramer}, {Kramer}, {Krichbaum}, {Kuo}, {Bella},
  {Lauer}, {Lee}, {Lee}, {Leung}, {Levis}, {Li}, {Lico}, {Lindahl},
  {Lindqvist}, {Lisakov}, {Liu}, {Liu}, {Liuzzo}, {Lo}, {Lobanov}, {Loinard},
  {Lonsdale}, {Lu}, {Mao}, {Marchili}, {Markoff}, {Marrone}, {Marscher},
  {Mart{\'\i}-Vidal}, {Matsushita}, {Matthews}, {Medeiros}, {Menten},
  {Michalik}, {Mizuno}, {Mizuno}, {Moran}, {Moriyama}, {Moscibrodzka},
  {M{\"u}ller}, {Mus}, {Musoke}, {Myserlis}, {Nadolski}, {Nagai}, {Nagar},
  {Nakamura}, {Narayan}, {Narayanan}, {Natarajan}, {Nathanail}, {Fuentes},
  {Neilsen}, {Neri}, {Ni}, {Noutsos}, {Nowak}, {Oh}, {Okino}, {Olivares},
  {Ortiz-Le{\'o}n}, {Oyama}, {{\"O}zel}, {Palumbo}, {Paraschos}, {Park},
  {Parsons}, {Patel}, {Pen}, {Pesce}, {Pi{\'e}tu}, {Plambeck}, {PopStefanija},
  {Porth}, {P{\"o}tzl}, {Prather}, {Preciado-L{\'o}pez}, {Psaltis}, {Pu},
  {Ramakrishnan}, {Rao}, {Rawlings}, {Raymond}, {Rezzolla}, {Ricarte},
  {Ripperda}, {Roelofs}, {Rogers}, {Ros}, {Romero-Ca{\~n}izales},
  {Roshanineshat}, {Rottmann}, {Roy}, {Ruiz}, {Ruszczyk}, {Rygl},
  {S{\'a}nchez}, {S{\'a}nchez-Arg{\"u}elles}, {S{\'a}nchez-Portal}, {Sasada},
  {Satapathy}, {Savolainen}, {Schloerb}, {Schonfeld}, {Schuster}, {Shao},
  {Shen}, {Small}, {Sohn}, {SooHoo}, {Souccar}, {Sun}, {Tazaki}, {Tetarenko},
  {Tiede}, {Tilanus}, {Titus}, {Torne}, {Traianou}, {Trent}, {Trippe}, {Turk},
  {van Bemmel}, {van Langevelde}, {van Rossum}, {Vos}, {Wagner},
  {Ward-Thompson}, {Wardle}, {Weintroub}, {Wex}, {Wharton}, {Wielgus}, {Wiik},
  {Witzel}, {Wondrak}, {Wong}, {Wu}, {Yamaguchi}, {Yoon}, {Young}, {Young},
  {Younsi}, {Yuan}, {Yuan}, {Zensus}, {Zhang}, {Zhao}, \&
  {Zhao}}]{EHT_SgrA_PaperVI}
---. 2022{\natexlab{f}},
  \href{http://dx.doi.org/10.3847/2041-8213/ac6756}{\JournalTitle{\apjl}, 930,
  L17}

\bibitem[{{Falla} \& {Floyd}(2002)}]{Falla2002}
{Falla}, D.~F., \& {Floyd}, M.~J. 2002,
  \href{https://ui.adsabs.harvard.edu/abs/2002EJPh...23...69F/abstract}{\JournalTitle{European
  Journal of Physics}, 23, 69}

\bibitem[{{Fromm} {et~al.}(2022){Fromm}, {Cruz-Osorio}, {Mizuno}, {Nathanail},
  {Younsi}, {Porth}, {Olivares}, {Davelaar}, {Falcke}, {Kramer}, \&
  {Rezzolla}}]{Fromm2022}
{Fromm}, C.~M., {Cruz-Osorio}, A., {Mizuno}, Y., {et~al.} 2022,
  \href{http://dx.doi.org/10.1051/0004-6361/202142295}{\JournalTitle{\aap},
  660, A107}

\bibitem[{{Georgiev} {et~al.}(2022){Georgiev}, Pesce, Broderick, Wong,
  {et~al.}}]{Georgiev2022}
{Georgiev}, B., Pesce, D.~W., Broderick, A.~E., Wong, G.~N., {et~al.} 2022,
  \href{http://dx.doi.org/10.3847/2041-8213/ac65eb}{\JournalTitle{\apj}, 930,
  0}

\bibitem[{{Gold} {et~al.}(2020){Gold}, {Broderick}, {Younsi}, {Fromm},
  {Gammie}, {Mo{\'s}cibrodzka}, {Pu}, {Bronzwaer}, {Davelaar}, {Dexter},
  {Ball}, {Chan}, {Kawashima}, {Mizuno}, {Ripperda}, {Akiyama}, {Alberdi},
  {Alef}, {Asada}, {Azulay}, {Baczko}, {Balokovi{\'c}}, {Barrett}, {Bintley},
  {Blackburn}, {Boland}, {Bouman}, {Bower}, {Bremer}, {Brinkerink},
  {Brissenden}, {Britzen}, {Broguiere}, {Byun}, {Carlstrom}, {Chael},
  {Chatterjee}, {Chatterjee}, {Chen}, {Chen}, {Cho}, {Christian}, {Conway},
  {Cordes}, {Crew}, {Cui}, {De Laurentis}, {Deane}, {Dempsey}, {Desvignes},
  {Doeleman}, {Eatough}, {Falcke}, {Fish}, {Fomalont}, {Fraga-Encinas},
  {Freeman}, {Friberg}, {G{\'o}mez}, {Galison}, {Garc{\'\i}a}, {Gentaz},
  {Georgiev}, {Goddi}, {Gu}, {Gurwell}, {Hada}, {Hecht}, {Hesper}, {Ho}, {Ho},
  {Honma}, {Huang}, {Huang}, {Hughes}, {Inoue}, {Issaoun}, {James}, {Jannuzi},
  {Janssen}, {Jeter}, {Jiang}, {Jimenez-Rosales}, {Johnson}, {Jorstad}, {Jung},
  {Karami}, {Karuppusamy}, {Keating}, {Kettenis}, {Kim}, {Kim}, {Kim}, {Kino},
  {Koay}, {Koch}, {Koyama}, {Kramer}, {Kramer}, {Krichbaum}, {Kuo}, {Lauer},
  {Lee}, {Li}, {Li}, {Lico}, {Lindqvist}, {Liu}, {Liuzzo}, {Lo}, {Lobanov},
  {Loinard}, {Lonsdale}, {Lu}, {MacDonald}, {Markoff}, {Mao}, {Marrone},
  {Marscher}, {Mart{\'\i}-Vidal}, {Matsushita}, {Matthews}, {Medeiros},
  {Menten}, {Mizuno}, {Moran}, {Moriyama}, {M{\"u}ller}, {Nagai}, {Nakamura},
  {Nagar}, {Narayan}, {Narayanan}, {Natarajan}, {Neri}, {Ni}, {Noutsos},
  {Okino}, {Ortiz-Le{\'o}n}, {Oyama}, {{\"O}zel}, {Palumbo}, {Park}, {Patel},
  {Pen}, {Pesce}, {Plambeck}, {Pi{\'e}tu}, {PopStefanija}, {Porth},
  {Preciado-L{\'o}pez}, {Psaltis}, {Ramakrishnan}, {Rao}, {Rawlings},
  {Raymond}, {Rezzolla}, {Roelofs}, {Rogers}, {Ros}, {Rose}, {Roshanineshat},
  {Rottmann}, {Roy}, {Ruszczyk}, {Rygl}, {S{\'a}nchez},
  {S{\'a}nchez-Arguelles}, {Sasada}, {Savolainen}, {Schuster}, {Schloerb},
  {Shao}, {Shen}, {Small}, {Sohn}, {SooHoo}, {Tiede}, {Tazaki}, {Tilanus},
  {Titus}, {Toma}, {Torne}, {Trent}, {Traianou}, {Trippe}, {Tsuda}, {van
  Langevelde}, {van Bemmel}, {van Rossum}, {Wagner}, {Wardle}, {Wex},
  {Weintroub}, {Wharton}, {Wielgus}, {Wong}, {Wu}, {Yoon}, {Young}, {Young},
  {Yuan}, {Yuan}, {Zensus}, {Zhao}, {Zhao}, {Zhu}, \& {Event Horizon Telescope
  Collaboration}}]{Gold2020}
{Gold}, R., {Broderick}, A.~E., {Younsi}, Z., {et~al.} 2020,
  \href{http://dx.doi.org/10.3847/1538-4357/ab96c6}{\JournalTitle{\apj}, 897,
  148}

\bibitem[{{Hada} {et~al.}(2013){Hada}, {Kino}, {Doi}, {Nagai}, {Honma},
  {Hagiwara}, {Giroletti}, {Giovannini}, \& {Kawaguchi}}]{Hada2013}
{Hada}, K., {Kino}, M., {Doi}, A., {et~al.} 2013,
  \href{http://dx.doi.org/10.1088/0004-637X/775/1/70}{\JournalTitle{\apj}, 775,
  70}

\bibitem[{{Kato} {et~al.}(2008){Kato}, {Fukue}, \& {Mineshige}}]{Kato2008}
{Kato}, S., {Fukue}, J., \& {Mineshige}, S. 2008, {Black-Hole Accretion Disks
  --- Towards a New Paradigm ---}

\bibitem[{{Kim} {et~al.}(2018){Kim}, {Krichbaum}, {Lu}, {Ros}, {Bach},
  {Bremer}, {de Vicente}, {Lindqvist}, \& {Zensus}}]{Kim2018a}
{Kim}, J.~Y., {Krichbaum}, T.~P., {Lu}, R.~S., {et~al.} 2018,
  \href{http://dx.doi.org/10.1051/0004-6361/201832921}{\JournalTitle{\aap},
  616, A188}

\bibitem[{{Kocherlakota} \& {Rezzolla}(2020)}]{Kocherlakota2020}
{Kocherlakota}, P., \& {Rezzolla}, L. 2020,
  \href{http://dx.doi.org/10.1103/PhysRevD.102.064058}{\JournalTitle{Phys. Rev.
  D}, 102, 064058}

\bibitem[{Liska {et~al.}(2018)Liska, Hesp, Tchekhovskoy, Ingram, van~der Klis,
  \& Markoff}]{Liska2018}
Liska, M., Hesp, C., Tchekhovskoy, A., {et~al.} 2018,
  \href{http://dx.doi.org/10.1093/mnrasl/slx174}{\JournalTitle{\mnras}, 474,
  L81}

\bibitem[{{Lu} {et~al.}(2023){Lu}, {Asada}, { Krichbaum}, {Park},
  {et~al.}}]{Lu2023}
{Lu}, R.-S., {Asada}, K., { Krichbaum}, T., {Park}, J., {et~al.} 2023,
  \href{https://ui.adsabs.harvard.edu/abs/2023Natur.616..686L/abstract}{\JournalTitle{Nature},
  616, 686}

\bibitem[{{MAGIC Collaboration} {et~al.}(2020){MAGIC Collaboration}, Acciari,
  Ansoldi, Antonelli, Arbet~Engels, Arcaro, Baack, Babić, Banerjee, Bangale,
  Barres~de Almeida, Barrio, Becerra~González, Bednarek, Bellizzi, Bernardini,
  Berti, Besenrieder, Bhattacharyya, \& Bigongiari}]{MAGIC2020}
{MAGIC Collaboration}, Acciari, V.~A., Ansoldi, S., {et~al.} 2020,
  \href{https://ui.adsabs.harvard.edu/abs/2020MNRAS.492.5354M/abstract}{\JournalTitle{\apj},
  492, 5354}

\bibitem[{{McKinney} {et~al.}(2012){McKinney}, {Tchekhovskoy}, \&
  {Blandford}}]{McKinney2012}
{McKinney}, J.~C., {Tchekhovskoy}, A., \& {Blandford}, R.~D. 2012,
  \href{http://dx.doi.org/10.1111/j.1365-2966.2012.21074.x}{\JournalTitle{\mnras},
  423, 3083}

\bibitem[{Mueller \& Lobanov(2022)}]{Muller2022}
Mueller, H., \& Lobanov, A.~P. 2022,
  \href{https://www.aanda.org/articles/aa/full_html/2022/10/aa43244-22/aa43244-22.html}{\JournalTitle{Astronomical
  Journal}, 666, 19}

\bibitem[{{Narayan} {et~al.}(2022){Narayan}, {Chael}, {Koushik}, {Ricarte},
  {et~al.}}]{Narayan2022}
{Narayan}, R., {Chael}, A., {Koushik}, C., {Ricarte}, A., {et~al.} 2022,
  \href{https://ui.adsabs.harvard.edu/abs/2022MNRAS.511.3795N/abstract}{\JournalTitle{\mnras},
  511, 3}

\bibitem[{Olivares {et~al.}(2019)Olivares, Porth, Davelaar, Most, Fromm,
  Mizuno, Younsi, \& Rezzolla}]{Olivares2019}
Olivares, H., Porth, O., Davelaar, J., {et~al.} 2019,
  \href{http://dx.doi.org/10.1051/0004-6361/201935559}{\JournalTitle{\aap},
  629, A61}

\bibitem[{Pandya {et~al.}(2016)Pandya, Zhang, Chandra, Gammie,
  {et~al.}}]{Pandya2016}
Pandya, A., Zhang, Z., Chandra, M., Gammie, C.~F., {et~al.} 2016,
  \href{https://ui.adsabs.harvard.edu/abs/2016ApJ...822...34P/abstract}{\JournalTitle{\apj},
  822, 34}

\bibitem[{{Porth} {et~al.}(2017){Porth}, {Olivares}, {Mizuno}, {Younsi},
  {Rezzolla}, {Moscibrodzka}, {Falcke}, \& {Kramer}}]{Porth2017}
{Porth}, O., {Olivares}, H., {Mizuno}, Y., {et~al.} 2017,
  \href{http://dx.doi.org/10.1186/s40668-017-0020-2}{\JournalTitle{Computational
  Astrophysics and Cosmology}, 4, 1}

\bibitem[{{Porth} {et~al.}(2019){Porth}, {Chatterjee}, {Narayan}, {Gammie},
  {Mizuno}, {Anninos}, {Baker}, {Bugli}, {Chan}, {Davelaar}, {Del Zanna},
  {Etienne}, {Fragile}, {Kelly}, {Liska}, {Markoff}, {McKinney}, {Mishra},
  {Noble}, {Olivares}, {Prather}, {Rezzolla}, {Ryan}, {Stone}, {Tomei},
  {White}, {Younsi}, {Akiyama}, {Alberdi}, {Alef}, {Asada}, {Azulay}, {Baczko},
  {Ball}, {Balokovi{\'c}}, {Barrett}, {Bintley}, {Blackburn}, {Boland},
  {Bouman}, {Bower}, {Bremer}, {Brinkerink}, {Brissenden}, {Britzen},
  {Broderick}, {Broguiere}, {Bronzwaer}, {Byun}, {Carlstrom}, {Chael},
  {Chatterjee}, {Chen}, {Chen}, {Cho}, {Christian}, {Conway}, {Cordes},
  {Geoffrey}, {Crew}, {Cui}, {De Laurentis}, {Deane}, {Dempsey}, {Desvignes},
  {Doeleman}, {Eatough}, {Falcke}, {Fish}, {Fomalont}, {Fraga-Encinas},
  {Freeman}, {Friberg}, {Fromm}, {G{\'o}mez}, {Galison}, {Garc{\'\i}a},
  {Gentaz}, {Georgiev}, {Goddi}, {Gold}, {Gu}, {Gurwell}, {Hada}, {Hecht},
  {Hesper}, {Ho}, {Ho}, {Honma}, {Huang}, {Huang}, {Hughes}, {Ikeda}, {Inoue},
  {Issaoun}, {James}, {Jannuzi}, {Janssen}, {Jeter}, {Jiang}, {Johnson},
  {Jorstad}, {Jung}, {Karami}, {Karuppusamy}, {Kawashima}, {Keating},
  {Kettenis}, {Kim}, {Kim}, {Kim}, {Kino}, {Koay}, {Patrick}, {Koch}, {Koyama},
  {Kramer}, {Kramer}, {Krichbaum}, {Kuo}, {Lauer}, {Lee}, {Li}, {Li},
  {Lindqvist}, {Liu}, {Liuzzo}, {Lo}, {Lobanov}, {Loinard}, {Lonsdale}, {Lu},
  {MacDonald}, {Mao}, {Marrone}, {Marscher}, {Mart{\'\i}-Vidal}, {Matsushita},
  {Matthews}, {Medeiros}, {Menten}, {Mizuno}, {Moran}, {Moriyama},
  {Moscibrodzka}, {M{\"u}ller}, {Nagai}, {Nagar}, {Nakamura}, {Narayanan},
  {Natarajan}, {Neri}, {Ni}, {Noutsos}, {Okino}, {Oyama}, {{\"O}zel},
  {Palumbo}, {Patel}, {Pen}, {Pesce}, {Pi{\'e}tu}, {Plambeck}, {PopStefanija},
  {Preciado-L{\'o}pez}, {Psaltis}, {Pu}, {Ramakrishnan}, {Rao}, {Rawlings},
  {Raymond}, {Ripperda}, {Roelofs}, {Rogers}, {Ros}, {Rose}, {Roshanineshat},
  {Rottmann}, {Roy}, {Ruszczyk}, {Rygl}, {S{\'a}nchez},
  {S{\'a}nchez-Arguelles}, {Sasada}, {Savolainen}, {Schloerb}, {Schuster},
  {Shao}, {Shen}, {Small}, {Sohn}, {SooHoo}, {Tazaki}, {Tiede}, {Tilanus},
  {Titus}, {Toma}, {Torne}, {Trent}, {Trippe}, {Tsuda}, {van Bemmel}, {van
  Langevelde}, {van Rossum}, {Wagner}, {Wardle}, {Weintroub}, {Wex}, {Wharton},
  {Wielgus}, {Wong}, {Wu}, {Young}, {Young}, {Yuan}, {Yuan}, {Zensus}, {Zhao},
  {Zhao}, {Zhu}, \& {Event Horizon Telescope Collaboration}}]{Porth2019}
{Porth}, O., {Chatterjee}, K., {Narayan}, R., {et~al.} 2019,
  \href{http://dx.doi.org/10.3847/1538-4365/ab29fd}{\JournalTitle{Astrophys. J.
  Supp.}, 243, 26}

\bibitem[{{Raymond} {et~al.}(2021){Raymond}, {Palumbo}, {Paine}, {Blackburn},
  {C{\'o}rdova Rosado}, {Doeleman}, {Farah}, {Johnson}, {Roelofs}, {Tilanus},
  \& {Weintroub}}]{Raymond2021}
{Raymond}, A.~W., {Palumbo}, D., {Paine}, S.~N., {et~al.} 2021,
  \href{http://dx.doi.org/10.3847/1538-3881/abc3c3}{\JournalTitle{\apjs}, 253,
  5}

\bibitem[{{Reid} {et~al.}(1982){Reid}, Schmitt, Owen, Booth, Wilkinson,
  Shaffer, Johnston, \& Hardee}]{Reid1982}
{Reid}, M.~J., Schmitt, J. H. M.~M., Owen, F.~N., {et~al.} 1982,
  \href{https://ui.adsabs.harvard.edu/abs/1982ApJ...263..615R/abstract}{\JournalTitle{\apj},
  263, 615}

\bibitem[{Reynolds \& Miller(2009)}]{Reynolds2009}
Reynolds, C.~S., \& Miller, M.~C. 2009,
  \href{https://ui.adsabs.harvard.edu/abs/2009ApJ...692..869R/abstract}{\JournalTitle{\apj},
  692}

\bibitem[{{Rezzolla} {et~al.}(2003){Rezzolla}, {Yoshida}, {Maccarone}, \&
  {Zanotti}}]{Rezzolla2003}
{Rezzolla}, L., {Yoshida}, S., {Maccarone}, T.~J., \& {Zanotti}, O. 2003,
  \href{https://ui.adsabs.harvard.edu/abs/2003MNRAS.344L..37R/abstract}{\JournalTitle{\mnras},
  344, L37}

\bibitem[{{Ricarte} {et~al.}(2023){Ricarte}, Tiede, Emami, Tamar,
  {et~al.}}]{Ricarte2022}
{Ricarte}, A., Tiede, P., Emami, R., Tamar, A., {et~al.} 2023,
  \href{https://ui.adsabs.harvard.edu/abs/2023Galax..11....6R/abstract}{\JournalTitle{Galaxies},
  11, 6}

\bibitem[{{Roelofs} {et~al.}(2023){Roelofs}, Blackburn, Lindahl, Doeleman,
  {et~al.}}]{Roelofs2023}
{Roelofs}, F., Blackburn, L., Lindahl, G., Doeleman, S., {et~al.} 2023,
  \href{https://ui.adsabs.harvard.edu/abs/2023Galax..11...12R/abstract}{\JournalTitle{Galaxies},
  11, 1}

\bibitem[{{Smith} {et~al.}(2021){Smith}, {Tandon}, \& {Wagoner}}]{Smith2021}
{Smith}, K.~L., {Tandon}, C.~R., \& {Wagoner}, R.~V. 2021,
  \href{https://ui.adsabs.harvard.edu/abs/2021ApJ...906...92S/abstract}{\JournalTitle{\apj},
  906, 8}

\bibitem[{{Snios} {et~al.}(2019){Snios}, Nulsen, Kraft, Cheung, Meyer, Forman,
  Jones, \& Murray}]{Snios2019}
{Snios}, B., Nulsen, P. E.~J., Kraft, R.~P., {et~al.} 2019,
  \href{https://ui.adsabs.harvard.edu/abs/2019ApJ...879....8S/abstract}{\JournalTitle{\apj},
  879, 8, 9}

\bibitem[{{Takahashi}(2004)}]{Takahashi2004}
{Takahashi}, R. 2004,
  \href{https://ui.adsabs.harvard.edu/abs/2004ApJ...611..996T/abstract}{\JournalTitle{Astrophys.
  J.}, 611, 996}

\bibitem[{{Tchekhovskoy} \& {McKinney}(2012)}]{Tchekhovskoy2012}
{Tchekhovskoy}, A., \& {McKinney}, J.~C. 2012,
  \href{http://dx.doi.org/10.1111/j.1745-3933.2012.01256.x}{\JournalTitle{\mnras},
  423, L55}

\bibitem[{{Thompson} {et~al.}(2017){Thompson}, {Moran}, \&
  {Swenson}}]{Thompson2017}
{Thompson}, A.~R., {Moran}, J.~M., \& {Swenson}, Jr., G.~W. 2017,
  {Interferometry and Synthesis in Radio Astronomy, 3rd Edition}

\bibitem[{Vagnozzi {et~al.}(2022)Vagnozzi, Roy, Tsai, Visinelli,
  {et~al.}}]{Vagnozzi2022}
Vagnozzi, S., Roy, R., Tsai, Y.-D., Visinelli, L., {et~al.} 2022,
  \href{https://ui.adsabs.harvard.edu/abs/2022arXiv220507787V/abstract}{\JournalTitle{arXiv:2205.07787}}

\bibitem[{{Xiao}(2006)}]{Xiao2006}
{Xiao}, F. 2006,
  \href{https://ui.adsabs.harvard.edu/abs/2023Natur.616..686L/abstract}{\JournalTitle{Plasma
  Physics and Controlled Fusion}, 48, 203}

\bibitem[{{Younsi} {et~al.}(2020){Younsi}, {Porth}, {Mizuno}, {Fromm}, \&
  {Olivares}}]{Younsi2020}
{Younsi}, Z., {Porth}, O., {Mizuno}, Y., {Fromm}, C.~M., \& {Olivares}, H.
  2020, \href{http://dx.doi.org/10.1017/S1743921318007263}{in Perseus in
  Sicily: From Black Hole to Cluster Outskirts, ed. K.~{Asada}, E.~{de Gouveia
  Dal Pino}, M.~{Giroletti}, H.~{Nagai}, \& R.~{Nemmen}, Vol. 342}, 9

\bibitem[{{Younsi} {et~al.}(2023){Younsi}, {Psaltis}, \&
  {{\"O}zel}}]{Younsi2023}
{Younsi}, Z., {Psaltis}, D., \& {{\"O}zel}, F. 2023,
  \href{https://ui.adsabs.harvard.edu/abs/2023ApJ...942...47Y/abstract}{\JournalTitle{Astron.
  Astrophys.}, 942, 23}

\bibitem[{{Younsi} {et~al.}(2012){Younsi}, {Wu}, \& {Fuerst}}]{Younsi2012}
{Younsi}, Z., {Wu}, K., \& {Fuerst}, S.~V. 2012,
  \href{http://dx.doi.org/10.1051/0004-6361/201219599}{\JournalTitle{\aap},
  545, A13}

\end{thebibliography}
\end{document}